\documentclass[preprint]{imsart}
\RequirePackage[OT1]{fontenc}
\RequirePackage{amsthm,amsmath}
\RequirePackage{natbib}
\RequirePackage[pdfencoding=auto, psdextra,  colorlinks,linkcolor=blue,citecolor=blue,urlcolor=blue,filecolor=blue]{hyperref}
\usepackage{a4wide,graphicx,color}
\usepackage{epstopdf}
\usepackage[11pt]{moresize}
\usepackage[font=footnotesize,caption=false]{subfig}
\usepackage{bookmark}
\newcommand{\vect}[1]{\boldsymbol{#1}}
\usepackage{array}
\usepackage{calc}
\usepackage{booktabs}
\usepackage{longtable}
\usepackage{tabularx}
\usepackage{multirow}
\usepackage[table]{xcolor}
\usepackage{mathrsfs}
\usepackage{amsfonts}
\usepackage{bbm}
\usepackage{lmodern}

\newcolumntype{C}{>{\centering\arraybackslash}X}
\newcolumntype{R}{>{\raggedright\arraybackslash}X}
\newcolumntype{L}{>{\raggedleft\arraybackslash}X}
\usepackage{blkarray}
\usepackage{pdflscape}
\usepackage{longtable}
\usepackage{siunitx}

\startlocaldefs
\numberwithin{equation}{section}
\theoremstyle{plain}
\newtheorem{proposition}{Proposition}
\endlocaldefs

\begin{document}

\begin{frontmatter}
\title{\Large{A Bayesian Covariance Graphical And Latent Position Model For Multivariate Financial Time Series}}
\runtitle{Bayesian Covariance Graphical And Latent Position Model}

\begin{aug}
\author{\fnms{Daniel F.} \snm{Ahelegbey}\ead[label=e1]{dfkahey@bu.edu}},
\author{\fnms{Luis} \snm{Carvalho}\ead[label=e2]{lecarval@math.bu.edu}}
\and
\author{\fnms{Eric} \snm{Kolaczyk}\ead[label=e3]{kolaczyk@bu.edu}}

\runauthor{D. F. Ahelegbey et al.}

\affiliation{Boston University}

\address{Department of Mathematics and Statistics, Boston University, Boston, MA, 02215 USA}
\end{aug}

\end{frontmatter}

\vspace{-5mm}

\noindent \textbf{Abstract:}
Current understanding holds that financial contagion is driven mainly by the system-wide interconnectedness of institutions. 
A distinction has been made between systematic and idiosyncratic channels of contagion, with shocks transmitted through the 
latter expected to be substantially more likely to lead to a systemic crisis than through the former. Idiosyncratic connectivity is 
thought to be driven not simply by obviously shared characteristics among institutions, but more by latent characteristics that 
lead to the holding of related securities. We develop a graphical model for multivariate financial time series with interest in 
uncovering the latent positions of nodes in a network intended to capture idiosyncratic relationships. We propose a hierarchical 
model consisting of a VAR, a covariance graphical model (CGM) and a latent position model (LPM). The VAR enables us to extract 
useful information on the idiosyncratic components, which are used by the CGM to model the network and the LPM uncovers the spatial 
position of the nodes. We also develop a Markov chain Monte Carlo algorithm that 
iterates between sampling parameters of the CGM and the LPM, using samples from the latter to update prior information for covariance 
graph selection. We show empirically that modeling the idiosyncratic channel of contagion using our approach can relate latent 
institutional features to systemic vulnerabilities prior to a crisis.

\vspace{5mm}
\noindent {\bf Keywords:} Bayesian inference, Covariance graph model, Idiosyncratic Contagion Channels, Latent Space Models,  Systemic Risk, VAR

\section{Introduction}

We approach the problem of modeling contagion by formulating a hierarchical
model with substructures consisting of a vector autoregressive (VAR) model, a
covariance graphical model (CGM)  and a latent position model (LPM). We refer
to our model approach as the Bayesian covariance graph and latent positions
model (BCGLPM).  The VAR framework is employed to allow us to extract useful
information about firm-level (idiosyncratic) factors from stock volatilities.
The CGM then uses this information to model the network of exposures as a
bi-directed structure determined by zeros in the covariance matrix of the
idiosyncratic shocks. The next step of the model adopts the LPM to uncover the
relative position of the firms in a latent space representing a global
financial market. This step is motivated by the idea that in many real world
settings, the actions and opportunities available to individuals depend on
their position in a social network. Also, financial firms tend to ``behave'' similarly in 
the market to form clusters. 
Therefore, learning these positions can
contribute to identifying ``behavioral'' clusters and improving our understanding of the pattern of interactions among
firms. For introduction to LPM and its variants,
see~\cite{Hoff_2008,Sarkar_2005,Sewell_2015,Kolaczyk_2014}. 

In recent years, networks have emerged as a critical tool for understanding
and managing contagion, and hence risk, in financial systems. Over the past
few years, especially after the 2008 financial crisis, there have been many
studies on financial  networks and their role in systemic risk. A major
finding emphasized by these studies is that financial contagion is mainly
driven by system-wide interconnectedness of institutions. The adoption of
networks is therefore critical to understand how fragile or robust the system
is to shocks; for example, to understand the threshold of connectedness beyond
which transmission of shocks can lead to a
crisis~\citep{Elliott_2014,Billio_2012,Diebold_2014,Acemoglu_2015,Ahelegbey_2016}.
Also, institutions such as the International Monetary Fund (IMF), the Bank for
International Settlements (BIS) and the Financial Stability Board (FSB)  have
adopted connectedness as one of the key factors for measuring emerging risks
and systemic vulnerabilities~\cite[see][]{IMF_2011,Moghadam_2010,Arregui_2013}.

\cite{Bernanke_2013} distinguishes triggers and vulnerabilities as the two key
factors that lead to a financial crisis. The triggers are the initial events
(losses, shocks) that affect many institutions in a financial market, and the
vulnerabilities are the pre-existing structural weaknesses of the system that
amplify these initial shocks. In the absence of vulnerabilities, the triggers
might produce sizable effects to a number of firms and investors, but would
generally not lead to a systemic breakdown. \cite{Tang_2010} showed that all
financial crises are alike and although the triggers may differ, the
vulnerabilities remain predominantly the same across systemic breakdowns. They
identified three potential channels for contagion effects: idiosyncratic,
market and country channels. In analyzing international contagion in the
banking industry during the crisis, \cite{Dungey_2015} also identified three
channels of contagion, namely, systematic, idiosyncratic and volatility
spillover. A key finding of the authors is that shocks transmitted via
idiosyncratic contagion increase the likelihood of a systemic crisis in the
domestic banking system by almost 37 percent, whereas increased exposure via
systematic contagion does not necessarily destabilize the domestic banking
system. 

We contribute to the wing of recent development in the application of VAR models for contagion
analysis by advocating for the decomposition of stock volatilities of financial firms into
the market (systematic) and firm-level (idiosyncratic) factors. This is to allow us monitor the
idiosyncratic channels of financial connectedness. We define systemic
vulnerabilities by connectedness beyond a tipping point that can amplify
losses in financial markets and cause systemic failure. Unlike other existing
approaches, we focus on modeling the position of firms in a latent space and
how these positions drive interconnectedness. 

Due to the levels of hierarchy in our model coupled with applications to
datasets with large number of variables, we are confronted with a number of
inferential challenges. Estimating the parameters of the model jointly is a
challenging inference problem and a computationally intensive exercise.
Following the literature on shrinkage methods~\citep[e.g.,][]{Tibshirani_1996,George_2008}, we advance a Bayesian approach
that incorporates relevant prior information to shrink the autoregressive
coefficients in the VAR. This allows us to focus on the inference of the
idiosyncratic channel of exposures and the latent firm positions - our primary
objective.  This is made feasible by integrating out the autoregressive
coefficients with respect to their prior distribution to obtain the marginal
likelihood function.  We build on a recently proposed prior distribution for
Gaussian graphical models by~\cite{Wang_2015} and the simulation of the
Bingham-von Mises-Fisher distribution for latent position models
by~\citep{Hoff_2009} to develop a Markov chain Monte Carlo algorithm. We
developed the MCMC to iterate between sampling parameters of the CGM and the
LPM, using samples from the latter to update prior information for covariance
graph selection. 

The contribution of this work are manifold. We contribute to the growing
literature on modeling firm-level idiosyncratic factors in the creation of
vulnerabilities for systemic risk. Secondly, we advance the literature
on covariance graph selection with automatic update of graph priors.  Thirdly,
we present an application of latent position models with the assumption
that the network structure is not known apriori and must be inferred from a
time series data. A fourth contribution is towards the potential application
of Procrustean analysis to financial times series, in interpreting the
inferred latent positions. For applications of the Procrustes approach for
investigating similarity of sets of spatial positions,
\cite[see][]{Dryden_2016,Cox_2000,Gower_2004,Wang_2010}.

We provide empirical applications of our approach to a high dimensional financial
time series. Since there are no existing approaches for joint inference of the
covariance graph and the latent nodal positions to the best of our knowledge, we compare the BCGLPM with
the stochastic search structure learning (SSSL) of~\cite{Wang_2015}. The result shows that by tracking firm-level
idiosyncratic information  and learning the position of the firms in a latent
space, BCGLPM performs better at uncovering the vulnerabilities of the global
financial system that existed between early-2004 to mid-2007, which amplified
the triggers of the financial crisis and the turmoil in its aftermath. The
Procrustean analysis of the network shows a more pronounced dissimilarity in
the nodal position of financial firms when tracking idiosyncratic factors, as
compared to market information, except for during the actual crisis periods.

The rest of the paper proceeds as follows. In Section~\ref{Model formulation},
we propose a hierarchical model, a discussion on the parameters and prior
specification and our Bayesian model
inference scheme. In Section~\ref{Simulation Experiments}, we provide an
illustration of BCGLPM on synthetic datasets and a comparison with alternative
approaches. Section~\ref{Financial Linkages And Nodal Positions} presents the
empirical financial application and results, and Section~\ref{Conclusion} concludes
the paper.
\section{Bayesian Covariance Graph and Latent Positions Model}
\label{Model formulation}

\subsection{Hierarchical Model Formulation}
We present the hierarchical model and a discussion of the
parameters and prior specification. The model consists of a vector
autoregression (VAR), a covariance graphical model (CGM) and a latent position
model (LPM). 

\begin{figure}[!ht]
\centering
\includegraphics[height=0.15\textheight]{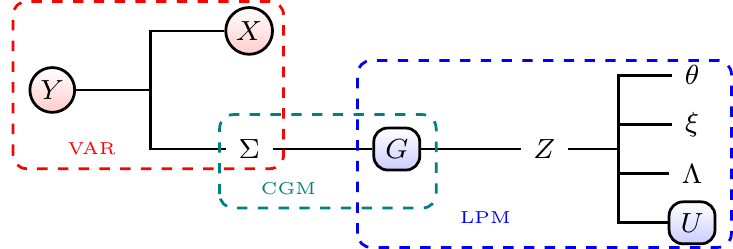} 
\caption{An illustration of the hierarchical model configuration. The red
circled variables represent the observed data and the blue rectangle variables
are our main parameters of interest.}
\label{Dynamic Latent Model}
\end{figure}

Figure~\ref{Dynamic Latent Model} shows an illustration of the model
configuration. The variables in the red-dashed rectangle constitute the VAR
sub-structure, the green-dashed rectangle constitute the CGM and the
blue-dashed rectangle represents the LPM structure. The variables in the
figure are defined as follows: $Y$ is a collection of the endogenous
variables, $X$ is a collection of the predictors composed of past observations
of the endogenous variables and market indicators, $\Sigma$ is a covariance
matrix of the idiosyncratic factors (shocks),  $G$ is a network of exposures
among the idiosyncratic factors, $Z$ is a similarity matrix whose elements are
constrained to be positive or negative depending on $G$, and $\{U, \Lambda,
\theta, \xi\}$ are parameters associated with an eigendecomposition of $Z$,
where $U$ is the latent coordinates matrix, $\Lambda$ is an eigenvalue matrix,
$\theta$ is a constant and $\xi$ is an error term. Of all the variables, our
primary objective is inference on the idiosyncratic shocks channel of
exposures, $G$, and the latent coordinates, $U$.

\subsubsection{VAR Model}
The idea for this model is to track information on the idiosyncratic factors
from observed time series. Let $Y_t$ be a vector of institutional financial
variables (e.g., log stock volatilities), $M_t$ is a vector of market
indicators representing the state of the economy, and $E_t$ is a vector 
of idiosyncratic factors. We model $Y_t$ and $E_t$ as follows:
\begin{align}\label{VARX}
  Y_t & = A_y Y_{t-1} + A_m M_{t-1} + E_t ~=  AX_t + E_t, \qquad E_t \sim  \mathcal{N}(0, \Sigma) \\
  \label{Error equation}
 E_t &= B E_t + \epsilon_t = (I_n - B)^{-1} \epsilon_t, \qquad \epsilon_t \sim \mathcal{N}(0, Q) 
\end{align} 
where $Y_t$ is an $n\times 1$ vector, $A = (A_y, A_m)$ is an $n \times k$
matrix of coefficients, $X_t = (Y_{t-1}, M_{t-1})'$ is a $k\times 1$ vector of
past observations, $E_t$ is an $n\times 1$ vector with $\Sigma$ as the
covariance matrix. From~\eqref{Error equation}, $\epsilon_t$ is an $n\times 1$
vector of error terms with a diagonal covariance matrix $Q$, $I_n$ is an
identity matrix of order $n$ and $B$ is an $n\times n$ matrix of coefficients
such that $B_{i,j}$ measures the contemporaneous exposure of firm $i$ to $j$
(or the contribution of $j$ to $i$). 

In this application, we assume the parameters $\{A_y, A_m, \Sigma, B, Q\}$ are fixed for a fixed window estimation. 
Thus, applying a moving window estimation allows us to estimate the parameters for the different windows. 
The autoregressive feature of the volatility of financial assets is supported
in the financial literature. It is well known that time series of financial
assets are often characterized by volatility clustering. That is, large
(small) changes in the prices of financial assets tend to be followed by large
(small) changes. Our choice of one lag for both exogenous and endogenous
variables is motivated by the persistence of volatilities, assuming a random
walk process on each variable with contributions from past information of
other variables and market indicators. From our experience with large VAR
models, most coefficients of higher lags tend to be concentrated around zero.
Despite these reasons, the model can easily be extended to cases of higher lag
orders or different lag orders on the exogenous and endogenous variables.  

In this application, we are particularly interested in identifying the
structure of exposures among the idiosyncratic variables. Thus, for each node
in the network, our goal is to identify the nodes that influence it and those
it influences.

\subsubsection{Covariance Graph Model (CGM)}
Let $Y= (Y_1',\ldots,Y_T')'$, $X = (X_1',\ldots,X_T')'$ and $E=
(E_1',\ldots,E_T')'$ be a collection of $Y_t$, $X_t$ and $E_t$ respectively
over a fixed window of length $T$. That is $Y$ and $E$ are matrices of
dimension $T\times n$, and $X$ is of dimension $T \times k$. Under the
assumption that $E$ is multivariate Gaussian distributed, there is a
correspondence between $B$, $Q$ and $\Sigma$. Given $\Sigma$, elements in $B$
and $Q$ can be obtained by 
\begin{align}\label{Correspondence beta sigma}
 B_{i,\pi_i} = \Sigma_{i,\pi_i}\Sigma_{\pi_i,\pi_i}^{-1} \qquad \text{and}
 \qquad Q_{ii} = \Sigma_{i,i} - \Sigma_{i,\pi_i}\Sigma_{\pi_i,\pi_i}^{-1}\Sigma_{i,\pi_i}'
\end{align}
where $B_{i,\pi_i}$ is the vector of coefficients from a univariate linear regression of $E_i$ on $E_{\pi_i}$ (the predictors of $E_i$) and $Q_{ii}$ is the covariance of $\epsilon_i$. 
We denote with $\pi_i$ the indexes of the predictors (parents) of node $i$, 
$\Sigma_{i,\pi_i}$ is the covariances between $E_i$ and $E_{\pi_i}$, and $\Sigma_{\pi_i,\pi_i}$ is the covariance among $E_{\pi_i}$. Here $E_i$ represents the $i$-th variable in $E$.
Given $B$ and $Q$, the covariance matrix $\Sigma$ can be obtained by
\begin{align}\label{Precision Matrix}
 \Sigma = (I - B)^{-1}Q(I - B)^{-1'} 
\end{align} 
where $A^{-1'}$ is the transpose of $A^{-1}$. 

The matrix $B$ encodes the relationship between the reduced-form errors in the sense that $B_{ij}\not=0$ if $E_j \to E_i$ and $0$ otherwise.  
From the correspondence between $B$ and $\Sigma$ in (\ref{Correspondence beta sigma}), the marginal independence relationships between any pair of variables in $E$ is such that, 
\begin{align}\label{Conditional Independence Relationship}
 E_i \leftrightarrow E_j \iff \Sigma_{i,j} \not= 0 \iff (B_{ij} \not= 0
 ~\text{and} ~ B_{ji} \not= 0)
\end{align}
We define a Gaussian bi-directed graph model, $G$, on $E$ which is determined
by zeros in $\Sigma$ \citep{Silva_2009}. This establishes a correspondence
between $\Sigma$ and $G$ such that $\Sigma_{ij}=0$ if and only if $G_{ij} =
G_{ji} =0$ and edges in $G$ corresponds to parameters in $\Sigma$.

\subsubsection{Latent Position Model (LPM)}
Modeling latent positions of agents have been successfully adopted in social
networks~\cite[see][]{Hoff_2008,Sewell_2015}. In this paper, we
advocate for the application of such models to learn the positions of
financial firms in a network instead of relying on assumptions of random or
predetermined positions (e.g., based on industry or firm size).
 
Let $U$ be an $n \times r$ matrix representing the coordinates of $n$ points
in an $r$-dimensional system. We use the notation $\vect{u}_i$ to denote the
$i$-th row of $U$ (i.e., the $i$-th node position), and $U_j$ to denote the
$n\times 1$ vector containing the $j$-th column of $U$.  For a convenient
visualization of the network structure, we set $r=2$, representing
two-dimensional spaces. We parameterize the $ij$-th entry of $G$ via a probit
mapping function following \cite{Hoff_2008, Hoff_2013}, given by 
\begin{align}\label{Probit Model 1}
  G_{ij} &=~ \vect{1}(Z_{ij} >0) \\
  \label{Probit Model 2}
  Z_{ij} &=~ \theta + \vect{u}_i' \Lambda \vect{u}_j + \xi_{ij}  =
  \theta + (U \Lambda U')_{ij} + \xi_{ij} 
\end{align}
where $\vect{1}(Z_{ij}>0)$ is the indicator function, i.e., unity if
$Z_{ij}>0$ and zero otherwise, $\theta$ is a constant, $(U \Lambda U')_{ij}$
is the $i$-th row and the $j$-th column of $(U \Lambda U')$, $\Lambda =
\text{diag} ~(\lambda_1, \lambda_2)$, is a diagonal matrix of eigen values,
and $\xi$ is a symmetric matrix of independent standard normal errors such
that $\xi_{ij}=\xi_{ji}$. The latent factor model above is an
eigen-decomposition of the graph $G$, where $U$ and $\Lambda$ are the
eigenvector and eigenvalue matrices, respectively.  From (\ref{Probit Model
2}), $Z$ can be explained as a similarity symmetric matrix which is normally
distributed, with mean $(\theta + U \Lambda U')$, and whose entries are
constrained to be positive or negative depending on $G$. 

As argued by~\cite{Hoff_2008}, the elements of $\Lambda$ help identify the
presence of homophily or stochastic equivalence---where nodes have similar
relational patterns with other nodes in the network. Suppose the latent
coordinates of node $i$ are similar to those of $j$, i.e., $\vect{u}_i \approx
\vect{u}_j$, and $U_{l,s}>0, ~l=i,j$ and $s=1,2$, and that the effect of
$\Lambda$ is such that $\lambda_i>0, ~i=1,2$.  Then there is a tendency for
nodes $i$ and $j$ to be connected. Thus, $\lambda_i>0$ indicates homophily.
Alternatively, if $\lambda_i<0, ~i=1,2$, then there is a tendency for nodes
$i$ and $j$ to be disconnected, although they share similar latent
coordinates. Thus, $\lambda_i<0$ indicates anti-homophily.

\subsection{Parameters and Prior Specification}
From the model discussed above, the parameters to estimate are $(A, \Sigma, G,
Z, U, \Lambda, \theta)$. In considering large numbers of variables with
relatively small sample size, we are confronted with the problem of
over-parameterization. Estimating parameters jointly is a challenging
inference problem and a computationally intensive exercise. 

We follow the literature on shrinkage methods~\citep{Tibshirani_1996,George_2008} by advancing a Bayesian paradigm
that incorporates relevant prior information to shrink the coefficients in $A$
while focusing on the inference of $G$ and $U$, our primary objective.
This is made feasible by integrating out $A$ with respect to its prior
distribution to obtain the marginal likelihood function.  The Bayesian
approach also includes simulation based approximations and model averaging
techniques to deal with parameter and model uncertainty.

For a fixed window of length $T$, we assume the idiosyncratic component, $E$,
follows a matrix-normal  distribution,
$E \sim \mathcal{MN}(\vect{0}, \Sigma, I_T)$, where $\vect{0}$ is a $T\times
n$ matrix of zeros, $\Sigma$ is the $n\times n$ row-specific covariance
matrix, and $I_T$ is the column-specific covariance matrix under the
assumption that the idiosyncratic terms are independent over time. 
The matrix form of~\eqref{VARX} can be expressed as $Y=XA' + E$. The
conditional distribution of $Y$ given $A$, $\Sigma$ and $X$ is given by,  
$Y|A, \Sigma \sim \mathcal{MN}(XA', \Sigma, I_T)$, whose likelihood function
is as follows: 
\begin{align}\label{Likelihood Function}
 P(Y| A,\Sigma) &= (2\pi)^{-\frac{nT}{2}}~|\Sigma|^{-\frac{T}{2}}
 ~\hbox{etr} \Big ( -\frac{1}{2} [\Sigma^{-1} (Y - XA')' (Y - XA')] \Big )
\end{align}
where $\hbox{etr}(\cdot)$ is the exponential of the standard trace function.

\subsubsection{Prior Distribution on \texorpdfstring{$A$}{A}}
We assume a natural-conjugate prior distribution, that is, a matrix-normal
conditional prior distribution of $A$ given $\Sigma$,
$A|\Sigma \sim \mathcal{MN}(\underline{A}, \Sigma, \Psi)$ with the following
density  
\begin{align} \label{Prior A}
 P(A|\Sigma) &= (2\pi)^{-\frac{1}{2}nk}~|\Psi|^{-\frac{1}{2}n}~|\Sigma|^{-\frac{1}{2}k}
 ~\hbox{etr} \Big ( -\frac{1}{2}  [\Sigma^{-1} (A-\underline{A})\Psi^{-1}
  (A-\underline{A})'] \Big )
\end{align}
where $\underline{A}$ is the prior mean of $A$, the row-specific prior
covariance matrix of $A$ is proportional to $\Sigma$, and the column-specific
prior covariance matrix of $A$ is proportional to $\Psi$. Note that
$\underline{A}, \Sigma$ and $\Psi$ are matrices of dimension $n\times k$,
$n\times n$ and $k\times k$ respectively. 

\begin{proposition}
\label{proposition 1}
Under the prior $P(A|\Sigma)$ in~\eqref{Prior A}, the likelihood $P(Y|
A,\Sigma)$ in~\eqref{Likelihood Function}, and with $S_{xx} = X'X + \Psi^{-1}$,
$S_{yx}=Y'X + \underline{A}\Psi^{-1}$,
$S_{yy} = Y'Y + \underline{A}\Psi^{-1}\underline{A}'$, and
$S_{y|x} = S_{yy} - S_{yx}S_{xx}^{- 1}S_{yx}'$, the marginal likelihood for any
covariance matrix $\Sigma$ is
\begin{align}\label{marginal likelihood}
  P(Y| \Sigma) & \propto ~|S_{y|x}|^{-\frac{n}{2}} ~\text{etr}
  \big ( -\frac{1}{2}  \Sigma^{-1} S_{y|x} \big ) \enskip .
\end{align}
\end{proposition}

\noindent  \textit{Proof}. Combining $P(Y| A,\Sigma)$ in~\eqref{Likelihood
Function} and $P(A|\Sigma)$ in~\eqref{Prior A}, the marginalization of $A$
with respect to its matrix-normal  conditional prior distribution is as
follows: 
\begin{align}\label{Marginalization}
 P(Y| \Sigma) & =  \int_A P(Y|A,\Sigma) ~  P(A|\Sigma) ~dA \nonumber \\ 
 & \propto  \int_A \hbox{etr} \Big ( -\frac{1}{2}  \Sigma^{-1} \Big [ A(X'X + \Psi^{-1} )A' -  2(Y'X + \underline{A}\Psi^{-1})A' + (Y'Y + \underline{A}\Psi^{-1}  \underline{A}') \Big ] \Big ) dA \nonumber \\ 
  & \propto  \int_A \hbox{etr} \Big ( -\frac{1}{2}  \Sigma^{-1} \Big [ A S_{xx} A' -  2S_{yx}A' + S_{yy} \Big ] \Big )  ~dA \nonumber \\
  & \propto \int_A \hbox{etr} \Big ( -\frac{1}{2}  \Sigma^{-1} \Big [ (A - S_{yx}S_{xx}^{- 1}) S_{xx} ( A - S_{yx}S_{xx}^{- 1})'  
  + S_{yy} - S_{yx}S_{xx}^{- 1}S_{yx}' \Big ] \Big ) dA.
\end{align}

From the expression above, the posterior distribution of $A$ is matrix-normal:
\begin{align}\label{posterior of A}
A|Y, X, \Sigma \sim \mathcal{MN}(S_{yx}S_{xx}^{- 1}, \Sigma, S_{xx}^{-1})
\end{align}
Substituting $\hat{A} = S_{yx}S_{xx}^{- 1}$ and
$S_{y|x} = S_{yy} - S_{yx}S_{xx}^{- 1}S_{yx}'$ into~\eqref{Marginalization},
\begin{align*}
  P(Y| \Sigma) & \propto  \hbox{etr} \Big ( -\frac{1}{2}  \Sigma^{-1} S_{y|x} \Big )   ~\int_A \hbox{etr} \Big ( -\frac{1}{2}  \Sigma^{-1} (A-\hat{A})S_{xx} (A-\hat{A})'  \Big )  ~dA \enskip .
\end{align*}
By definition
$\displaystyle \int_A \hbox{etr} \Big ( -\frac{1}{2}  \Sigma^{-1} (A-\hat{A})S_{xx}
 (A-\hat{A})'  \Big )  ~dA ~=
 (2\pi)^{\frac{nk}{2}}|S_{xx}|^{-\frac{n}{2}}|\Sigma|^{\frac{k}{2}}$ and 
\begin{align}\label{Marginalize A}
  P(Y| \Sigma) & = (2\pi)^{-\frac{nT}{2}} ~|\Psi|^{-\frac{n}{2}}~|S_{xx}|^{-\frac{n}{2}} ~|\Sigma|^{-\frac{T}{2}} ~
  \hbox{etr} \Big ( -\frac{1}{2}  \Sigma^{-1} S_{y|x} \Big ) \enskip .
\end{align}
%

\subsubsection{Prior Distribution on \texorpdfstring{$\Sigma$}{\Sigma} and
\texorpdfstring{$G$}{G}}
The common prior for $\Sigma$ is a $G$-inverse Wishart distribution. A general
problem of such distributions is that the normalizing constant is intractable.
Also, scalability to large graphical models is not feasible due to the
computational cost of approximating the normalizing
constant~\citep{Lenkoski_2011,Wang_2015}. 

Following the spike and slab priors of \cite{Wang_2015}, we construct a prior
distribution on $\Sigma$ by assuming an independent distribution on its
diagonals and off-diagonal elements.  We denote by $v_0$ and $v_1$  the
standard deviations of the spike and slab components, respectively. We define
$V$ to be an $n \times n$ symmetric matrix with ones on the diagonal, i.e.,
$V_{ii}=1$, and off-diagonals $V_{ij} = v_0^2$ if $G_{ij}=0$ and $V_{ij} =
v_1^2$ if $G_{ij}=1$, $v_1>v_0$.  The choice of values for $v_0$ and $v_1$ is
discussed in Section~\ref{Hyperparameters}. We assume the off-diagonals of
$\Sigma$ are normally distributed,
$\Sigma_{i,j}|G \sim \mathcal{N}(0,V_{ij})$, and the diagonals follows an
exponential distribution, $\Sigma_{i,i}|G \sim$ Exp$(\frac{1}{2}V_{ii})$. The
densities of the hierarchical prior distributions for $\Sigma|G$ and $G|Z$ are
given by
\begin{align}\label{Prior on Sigma}
 P(\Sigma|G) ~& = \prod_{i \not= j} \exp \Big ( -\frac{1}{2}
 V_{ij}^{-1}\Sigma_{i,j}^2 \Big )
 \prod_{i=1}^n \exp \Big (-\frac{1}{2} V_{ii} \Sigma_{i,i} \Big )
 \mathbf{1}(\Sigma \in \mathcal{S}_+(G)) \\
 \label{Prior on G}
 P(G|U, \Lambda, \theta) ~& =~ \prod_{i \not= j} \Gamma_{ij}^{G_{ij}}
 \big(1-\Gamma_{ij} \big)^{(1-G_{ij})}  \enskip ,
\end{align}
where $\mathcal{S}_+(G)$ is the space of symmetric positive definite matrices
with non-zero entries according to $G$ and $\Gamma_{ij}$ is the probability of
a link between nodes $i$ and $j$:
\begin{align}
~~ \Gamma_{ij}(U, \Lambda, \theta) ~=~ P(G_{ij}=1| U, \Lambda, \theta)~=~
P(Z_{ij} > 0 | U, \Lambda, \theta) ~=~ \Phi(\theta + (U \Lambda U')_{ij}),
\end{align}
where $\Phi$ is the cumulative density function of the standard normal
distribution.

\subsubsection{Distribution on (\texorpdfstring{$Z$}{Z},
\texorpdfstring{$\theta$}{\theta},  \texorpdfstring{$\Lambda$}{\Lambda},
\texorpdfstring{$\xi$}{\xi}, \texorpdfstring{$U$} U)} 
We assume the following prior distribution 
\begin{align}
Z | \theta, U, \Lambda &~ \sim ~ \mathcal{N}(\theta + U \Lambda U', \Sigma_Z) \\
 \theta & ~ \sim~ \mathcal{N}(\theta_0, \tau_{\theta}^2) \\
  U & ~\sim~ Unif (\mathcal{V}_r(\mathcal{R}^n)) \\
 \label{Prior lambda}
 (\lambda_1, \lambda_2) & ~ \overset{iid}{\sim}~ \mathcal{N}(0, \tau_{\lambda}^2) \\
  \label{Prior xi}
(\xi_{ij}=\xi_{ji}) & ~\overset{iid}{\sim}~ \mathcal{N}(0,1)
\end{align}
where $\theta_0$ and $\tau_{\theta}^2$ are the prior mean and variance of
$\theta$, and $\mathcal{V}_r(\mathcal{R}^n)$ is the Stiefel manifold of $r$
orthonormal vectors in $\mathcal{R}^n$.  We assume $U$ is uniformly
distributed on $\mathcal{V}_r(\mathcal{R}^n)$, such that $U'U=I_r$, where
$I_r$ is an identity matrix of order $r$. The prior distribution on the
elements of $\Lambda$ and $\xi$ is an independent normal distribution. The
covariance matrix $\Sigma_Z$ is such that the off-diagonals of $Z$ have unit
variances. 

\subsection{Bayesian Model Inference}

Given the data, $Y$, and the prior distributions on the parameters, 
we are particularly interested in the posterior inference of the graph structure, $G$, and the latent nodal positions, $U$. 
We proceed by performing the necessary posterior approximations through application of a Gibbs sampler that consists of the following steps:
\[
[ \Sigma | G, Y ], ~[ G | \Sigma, U, \Lambda, \theta ], ~ [Z | G, U, \Lambda, \theta ], ~ [ \theta | Z, U, \Lambda ], ~ [\Lambda | Z, U, \theta], ~\text{and} ~[U | Z, \Lambda, \theta].
\]
Note that the second Gibbs step is collapsed, that is, we marginalize $Z$. 
We sample $\{\Sigma, G\}$ following the results from  \cite{Wang_2015} and $\{Z, \theta, \Lambda, U \}$ following \cite{Hoff_2009}. 
See Appendix \ref{app} for detailed descriptions of the sampling approach of the parameters.
\section{Simulation Experiments}
\label{Simulation Experiments}

We evaluate the efficiency of our inference approach on simulated datasets from an $n$-node graphical model. 
We consider the following data generating process (DGP): 
\begin{align*}
 \textit{Lag-0 Setup}: ~&~ Y_t = E_t, \quad E_t \sim \mathcal{N}(0, \Sigma_G) \\
 \textit{Lag-1 Setup}: ~&~ Y_t = AY_{t-1} + E_t, \quad E_t \sim \mathcal{N}(0, \Sigma_G)
\end{align*}
where $\Sigma_G$ is a covariance matrix constrained by a sparse graph $G$, such that $\Sigma_{ij}=0$ if $G_{ij} =0$ and $\Sigma_{ij}\not=0$ if $G_{ij} =1$. 
The \textit{Lag-0 Setup} generates data from a regular Gaussian covariance graph model and the 
\textit{Lag-1 Setup} incorporates an AR(1) temporal dependence on the endogenous variables with a Gaussian covariance graph structure on the residuals.

We generate edges in $G$ from independent Bernoulli distributions with probability $0.2$. We construct $A$ to be a diagonal matrix. 
Following the random graph pattern of \cite{Wang_2012}, we construct $\Sigma_G = B_G +\delta I_n$, where $B_G$ is a symmetric matrix constrained by $G$. We generate $B_G$, $A$ 
and $\delta$ as follows:
\begin{align*}
 (B_G)_{ij} = \left\{
\begin{array}{ll}
 1  & \text{if} ~ i=j \\
 \beta_{ij}  & \text{if} ~ G_{ij}=1\\ 
 0 & \text{otherwise}
\end{array}
\right., 
&& 
 A_{ij} = \left\{
\begin{array}{ll}
\alpha_{i}  & \text{if} ~ i=j\\ 
 0 & \text{otherwise}
\end{array}
\right.,  && 
\delta = \frac{n \min(\delta_b) - \max(\delta_b)}{1-n}
\end{align*}
where $\delta_b$ is the eigenvalues of $B_G$. The values of $(\beta_{ij}= \beta_{ji})$ and $\alpha_{i}$ are randomly drawn from a uniform distribution on $(-0.9, -0.3) \cup (0.3, 0.9)$. 
For the \textit{Lag-1 Setup}, we initialize $Y_0 \sim \mathcal{N}(0, I_n)$.  
For each setup, we generate a dataset of dimension $n \in \{50, 100, 150\}$ and sample size $T \in \{2n, 10n\}$. We replicate the simulation and estimation exercise 10 times. 

\subsection{Setting Hyperparameters}
\label{Hyperparameters}
For the hyperparameters governing the distribution of $\theta$, i.e., ($\theta_0, \tau_{\theta}^2$), 
we notice that $\theta_0$ is positively related the prior edge probabilities, $\Gamma_{ij}$. Negative values of $\theta$ imply a lower or negative value for elements in $Z$, which leads to lower prior edge probabilities. 
We consider the problem of estimating a graphical model for large numbers of variables as one that can be approximated by a sparse structure. Thus, one would expect that the prior expectation of $\theta$ 
is negative. We set $\theta_0 = -0.5$ and $\tau_{\theta}^2 = 100$.

For the spike and slab parameters, we assume $v_0$ to be small, concentrated around zero, and $v_1 = h v_0$, a scaled version of $v_0$, 
where $h>0$ is large. By standardizing all datasets in our applications, we set   
$v_1=1$ and vary $(v_0, h) = (0.02, 50), (0.05, 20)$.

Following \cite{Hoff_2009}, we set $\tau_{\lambda}^2=n$. As argued by the author, this value reflects the variance of the eigenvalues of an $n\times n$ matrix of independent standard normal noise. 

We assume the prior expectation of $A$ is a zero matrix, $\underline{A}= \vect{0}$. We further assume 
the coefficients in $A$ are a-priori independent within and across equations. Thus, $\Psi = \eta^{-1} I_k$ is a diagonal matrix, where $\eta^{-1}$ is the prior variance 
and $k$  is the number of covariates in $X$. From our choice of $\underline{A}= \vect{0}$, the posterior expectation of $A$ in (\ref{posterior of A}) becomes 
$ \hat{A}'  =  (X'X + \eta I_k)^{- 1}X'Y$, which is the same as the ridge estimator with $\eta$ as the ridge parameter. 
We set $\eta = c_0 k$, where $c_0$ varies on a grid between $0.1$ to $10$. 
To determine the optimal choice of $c_0$, we divide the data into 80\% for estimation of $\hat{A}'$ and 20\% for point forecast evaluation. 
We obtain the mean squared forecast error (MSFE) of the point forecasts. 
We restrict our choice of $c_0$ to the grid where the first difference of the MSFE is less than a tolerance level, e.g., 0.01. This is to 
avoid over fitting the data as well as not shrinking all coefficients to zero.

\subsection{Competing Inference Approaches}
Since there are no existing approaches for joint inference of the graph and the latent nodal positions, it is reasonable to compare our 
approach with existing methods suitable for large covariance selection problems. To the best of our knowledge, the stochastic search structure learning (SSSL) method by \cite{Wang_2015} appear to be a suitable benchmark 
to compare our graph inference performance since it has been shown to be effective in dealing with large problems and complicated models. Following the suggestion by the author of that paper, we 
set $v_0=0.02, h=50$ and the hyper-parameter for the graph priors, $\pi=2/(n-1)$. 

We present two versions of our BCGLPM approach for the inference of the graph. We denote by BCGLPM(0) an application of the BCGLPM under the assumption 
that the data is generated from a contemporaneous dependence model. That is, BCGLPM(0) is a lag-0 version of our model.  Similarly, we denote by BCGLPM(1) the application of the BCGLPM under the assumption 
that the data is simulated from a VAR(1) model. 

The former assumes that the observed data follows a regular Gaussian covariance graphical model 
and the latter assumes a VAR structure with idiosyncratic components that follow a Gaussian covariance graph structure. In both cases, our goal is to infer the idiosyncratic graph structure and the latent 
nodal positions. To implement the BCGLPM(0) means setting $X$ to be a null matrix which leads to replacing $S_{y|x}=S_{yy} = Y'Y$ in (\ref{Marginalize A}). 
Since the BCGLPM(0) is closely related to the SSSL, the former is expected to be very competitive against the latter. 

\subsection{MCMC Diagnostics}
We run a total of 10,000 MCMC iterations for all competing methods with the first 3,000 as the burn-in sample.  
All computations were implemented in MATLAB through the Boston University Shared Computing Cluster.  
To monitor the mixing of the MCMC chain, we compute the negative log likelihood score given by 
\begin{align}\label{Log predictive density}
 -2 \log \mathcal{L}(Y|\cdot) ~ & = ~ -2\log P(Y| \Sigma, G)  -2\log P(Z|\theta, U, \Lambda) \nonumber \\
 ~ & = ~ T\log|\Sigma| + \text{tr} \big (S_{y|x}\Sigma^{-1}  \big )
 + \frac{1}{2}\text{tr} \big [ (E_z - U\Lambda U')'(E_z - U\Lambda U') \big ]
\enskip .
\end{align}
We use the negative log likelihood score to compute the potential scale reduction factor (PSRF) of \cite{Gelman_1992}.  
The chain is said to have converged if $PSRF \leq 1.2$. 

\subsection{Graph Predictive Evaluation}
To assess the performance of the graph structure inference, we compute the graph accuracy (ACC) and the AUC - the area under the receiver operator characteristic (ROC) curve. 
The AUC is a well-known measure of graph predictiveness. The higher the ACC and AUC the better the performance. 
We report the average $TP$ (true positives), $FP$ (false positives), $ACC$ and $AUC$ for each of the competing methods.  

\subsection{Results}
Table \ref{Summary of Model Performance} presents a summary of the comparison of the graph performance on the $100$ and $150$-node models with different sample sizes. 
For the different choice of $v_0$ and $h$, the result shows a more sparse structure for higher values of $v_0$. 
Since $v_0$ is positively related to the posterior edge inclusion probability, higher values of $v_0$ lead to a lower number of $TP$ and $FP$. 
The performance of BCGLPM(0) and BCGLPM(1) shows that ($v_0=0.02, h=50$) performs better at predicting the graph of the DGP than that of ($v_0=0.05, h=20$). 
Though the former seems dense, the $ACC$ and $AUC$ are higher than the latter. 

\begin{table}[!ht]
\scriptsize
\centering
\caption{Comparing inference performance on $100$ and $150$-node models with different sample sizes. 
The values of the $TP$ (true positives), $FP$ (false positives),  $ACC$ (accuracy), and $AUC$ (area under the ROC curve) are the average values across 10 simulations. 
Boldface values indicate the best choice for each metric. 
}
\begin{tabularx}{0.98\textwidth}{ C l  L   L L  L L  }
\toprule
  &   & SSSL     & \multicolumn{2}{c}{BCGLPM(0) }   & \multicolumn{2}{c}{BCGLPM(1) } \\ \cmidrule(r){3-3} 	\cmidrule(lr){4-5} \cmidrule(lr){6-7}
 $n$ & $(v_0, h)$  &  $(0.02, 50)$  &  $(0.02, 50)$ & $(0.05, 20)$ &  $(0.02, 50)$ & $(0.05, 20)$ \\ 
 \midrule
 \multicolumn{2}{c}{\textit{Lag-0 Setup}} &  \multicolumn{5}{c}{\textit{Sample Size} $~T=2n$} \\  
 \cmidrule(r){1-2}
 \multirow{4}{*}{100}
 &TP & 32.60 &  170.40 &  98.30 &  187.50 &  114.20  \\ 
 &FP & 1.00 &  17.90 &  4.20 &  46.60 &  15.10  \\ 
 &ACC & 80.64 &  \textbf{83.08} &  81.90 &  82.84 &  82.00  \\ 
 &AUC & 52.55 &  63.16 &  60.64 &  \textbf{63.59} &  61.01  \\ 
\addlinespace
 \multirow{4}{*}{150} 
 &TP & 63.70 &  320.90 &  123.70 &  373.30 &  148.30  \\ 
 &FP & 0.90 &  26.50 &  1.90 &  72.20 &  10.20  \\ 
 &ACC & 80.66 &  82.73 &  81.18 &  \textbf{82.79} &  81.33  \\ 
 &AUC & 52.64 &  61.63 &  56.57 &  \textbf{62.29} &  57.32  \\ 
\addlinespace
& &   \multicolumn{5}{c}{\textit{Sample Size}, $~T=10n$} \\  
\addlinespace
 \multirow{4}{*}{100}
 &TP & 356.80 &  634.00 &  285.70 &  633.70 &  290.40  \\ 
 &FP & 0.00 &  2.80 &  0.00 &  5.00 &  0.10  \\ 
 &ACC & 87.49 &  \textbf{93.03} &  86.05 &  92.98 &  86.14  \\ 
 &AUC & 67.00 &  \textbf{87.61} &  61.86 &  87.37 &  64.59  \\  
\addlinespace
 \multirow{4}{*}{150} 
 &TP & 507.70 &  1261.40 &  264.90 &  1254.90 &  268.80  \\ 
 &FP & 0.00 &  2.60 &  0.00 &  4.10 &  0.00  \\ 
 &ACC & 84.40 &  \textbf{91.12} &  82.23 &  91.05 &  82.26  \\ 
 &AUC & 58.77 &  \textbf{83.64} &  52.38 &  83.55 &  52.54  \\ 
  \midrule
 \multicolumn{2}{c}{\textit{Lag-1 Setup}} &  \multicolumn{5}{c}{\textit{Sample Size} $~T=2n$} \\  
 \cmidrule(r){1-2}
 \multirow{4}{*}{100}  
 &TP & 30.30 &  84.60 &  60.70 &  80.00 &  49.10  \\ 
 &FP & 26.00 &  88.10 &  49.80 &  21.60 &  9.00  \\ 
 &ACC & 80.08 &  79.92 &  80.21 &  \textbf{81.17} &  80.80  \\ 
 &AUC & 51.81 &  54.72 &  54.25 &  \textbf{56.57} &  55.40  \\ 
\addlinespace
 \multirow{4}{*}{150}  
 &TP & 51.20 &  172.90 &  89.90 &  154.10 &  60.80  \\ 
 &FP & 41.00 &  177.00 &  73.20 &  33.60 &  5.70  \\ 
 &ACC & 79.97 &  79.85 &  80.03 &  \textbf{80.96} &  80.38  \\ 
 &AUC & 51.38 &  54.26 &  53.06 &  \textbf{55.66} &  53.37  \\ 
\addlinespace
& &   \multicolumn{5}{c}{\textit{Sample Size}, $~T=10n$} \\  
\addlinespace
 \multirow{4}{*}{100}  
 &TP & 140.40 &  291.50 &  109.50 &  602.60 &  242.50  \\ 
 &FP & 11.50 &  57.50 &  9.00 &  3.20 &  0.00  \\ 
 &ACC & 82.43 &  84.55 &  81.85 &  \textbf{91.93} &  84.72  \\ 
 &AUC & 58.89 &  67.03 &  58.73 &  \textbf{85.47} &  60.81  \\
\addlinespace
 \multirow{4}{*}{150}  
 &TP & 199.40 &  525.30 &  116.10 &  1153.70 &  213.20  \\ 
 &FP & 12.10 &  96.50 &  7.30 &  1.80 &  0.00  \\ 
 &ACC & 81.91 &  84.07 &  81.20 &  \textbf{90.54} &  82.14  \\ 
 &AUC & 56.09 &  64.40 &  54.87 &  \textbf{82.09} &  52.00  \\ 
 \bottomrule
 \end{tabularx}
\label{Summary of Model Performance}
\end{table}

Now comparing the BCGLPM(0) and BCGLPM(1) under ($v_0=0.02, h=50$) with SSSL as the benchmark, we notice that overall, the BCGLPM approaches record more links and a higher $ACC$ and $AUC$ than the SSSL. 
Although the SSSL and the BCGLPM(0) follows a regular Gaussian covariance graphical model and a similar inference scheme, the results show a higher graph predictive performance of the latter over the former. 
Clearly, the idea of incorporating inference of the node positions and its application to update the graph priors contributes significantly to improving the network obtained by the BCGLPM(0) as 
compared to the fixed graph priors in SSSL. 

From the table, we also notice that though the graph inference accuracy of the methods reduces for relatively small sample sizes, 
the BCGLPM(1) is highly competitive against the others when fitting high dimensional Gaussian covariance graph models on data generated from a contemporaneous model or lag-0 process. 
Furthermore, the BCGLPM(1) outperforms the others when the DGP includes an autoregressive dependence (or lag-1) process. 
Overall, the result shows that by tracking the idiosyncratic factors from a time series of many variables and learning the latent nodal positions of these variables, the BCGLPM approach infers 
a more accurate structure of the idiosyncratic channel of exposures than the SSSL. 

\section{Volatility Connectedness And Nodal Position Analysis}
\label{Financial Linkages And Nodal Positions}

We analyze the stability of the volatility linkages and nodal positions in the financial sector of the US and European stock market. 
We focus on publicly traded financial institutions in the United States and Europe, consisting of three financial
sectors: banks (25), insurance companies (25), and real estate companies (20). Our sample period covers January 2002 to August 2014. The list of 
institutions also included four top US banks that have been acquired or bankrupted during the 2008 crisis: 
Bear Sterns, Countrywide Financial Corporation, Lehman Brothers and Merrill Lynch. We added 6 major 
global market indexes: S\&P 500, Dow Jones, Nasdaq Composite, Euro Stoxx 600, Euro Stoxx 50 and HangSeng index.

We obtain daily price indexes on the global market indicators from Yahoo finance and on the rest of the institutions from Datastream. 
We have in our sample a total of 150 institutions (includes the 4 acquired or bankrupted institutions) before 2008 and 146 after 2008. 
Countries represented by the different institutions in our data set include Austria, Belgium, Switzerland, Germany, Denmark, Spain, Finland, France, Greece, 
Ireland, Italy, the Netherlands, Norway, Sweden, the UK and the US. 
See Table \ref{Dataset 1} for detailed descriptions of the financial dataset.

To set up our model, we construct the stock volatilities as the squared returns of the daily price indexes given by \citep[see][]{Martens_2007}: 
\begin{equation}\label{volat}
RV_t = \Big [ 100 \Big (\log I_t-\log I_{t-1} \Big) \Big]^2
\end{equation}
where $I_t$ is the price index at the time $t$. We estimate the following model setups: 
\begin{align}
\label{lag_0}
 \textit{Lag-0 Setup}: ~&~ Y_t = E_t, \quad E_t \sim \mathcal{N}(0, \Sigma_G) \\
 \label{lag_1}
 \textit{Lag-1 Setup}: ~&~ Y_t = A_y Y_{t-1} + A_m M_{t-1} + E_t, \quad E_t \sim \mathcal{N}(0, \Sigma_G)
\end{align}
where $Y_t$ and $M_t$ denotes the volatilities of the firm-specific and market indicator variables, respectively. 
Similar to our simulation exercise, we model (\ref{lag_0}) with BCGLPM(0) and (\ref{lag_1}) with a BCGLPM(1) process. 
Again, we chose the SSSL method by \cite{Wang_2015} as a suitable benchmark to compare our graph inference.

Our empirical analysis is carried out using a yearly moving window estimation to allow inference 
on the stability of the connectedness structure and the similarity of nodal positions over the sample period. 
The first window considered is from January 2002 through December 2002 and the last window 
covers October 2013 to August 2014. 

\subsection{Network Density Analysis}
We define connectedness among institutions through the contemporaneous linkages among the idiosyncratic components captured by the covariance graph 
structure. We characterize (through numerical summaries) the time-varying nature of the interconnections by monitoring the network density. 

The network density is a simple aggregate index for the extent of interdependence. It is defined for each estimation window as the number of estimated links in the
network divided by the total number of possible edges. For $n$ number of firms, there are $n(n - 1)/2$ possible edges. 
The higher the network density the more the system-wide connectedness of financial firms, which indicates potential vulnerability of the system. Here the links in the network 
play a significant role in the transmission of financial risk or shocks \cite[see][]{Tang_2010, Billio_2012, Diebold_2014}. 

\begin{figure}[!ht]
\footnotesize
\centering
\subfloat[]{\includegraphics[height=0.25\textheight]{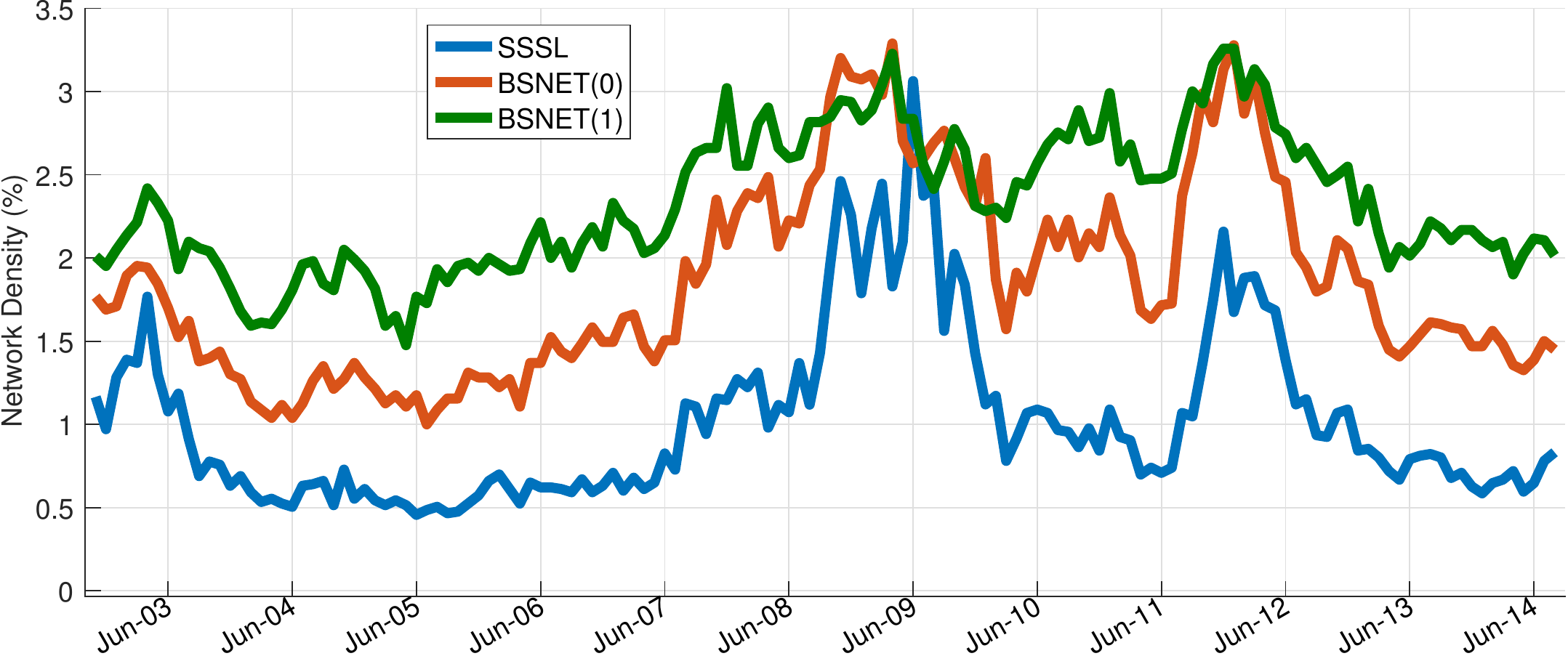} \label{Network Density}} \qquad 
\subfloat[]{\includegraphics[height=0.25\textheight]{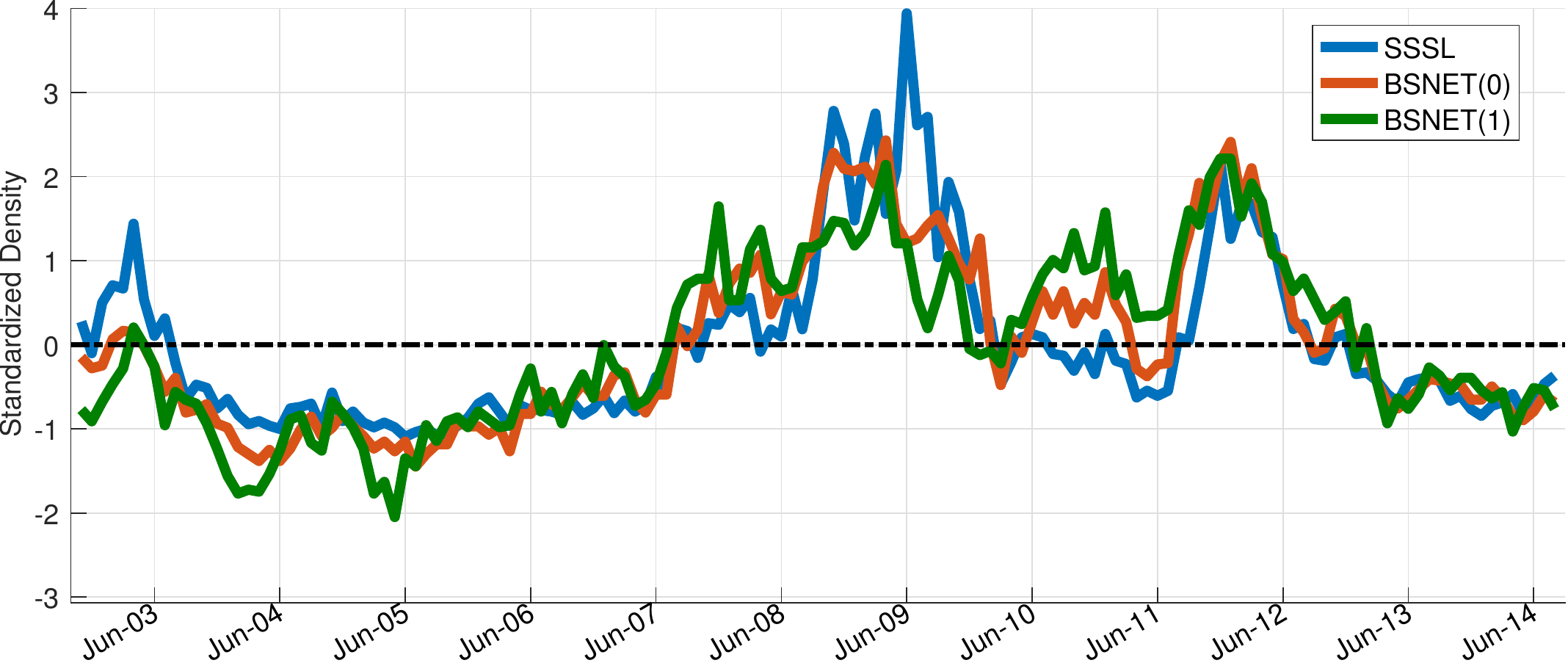} \label{Standardized Network Density}} 
\caption{(\ref{Network Density}) Network Densities and (\ref{Standardized Network Density}) Standardized Densities 
of the SSSL in blue, BCGLPM(0) in red and BCGLPM(1) in green, obtained from a one-year moving window estimation 
 from January, 2002 - August, 2014. The dashed zero-line is a natural choice of a threshold where negative densities suggest bearable connectedness
and positive densities indicate potential systemic vulnerability for shock amplification. }
\label{Financial Plots}
\end{figure}
 
\begin{table}[!ht]
\footnotesize
\centering
\caption{Comparing the critical periods with positive standardized network densities.} 
\begin{tabularx}{\textwidth}{LcR @{\extracolsep{\fill}} c LcR @{\extracolsep{\fill}} c LcR}
\toprule
 \multicolumn{3}{c}{SSSL }  && \multicolumn{3}{c}{BCGLPM(0) }   && \multicolumn{3}{c}{BCGLPM(1) } \\ \toprule
Dec-02 & -- &   Jul-03      &&      Feb-03 & -- &   Apr-03      &&        \\
Aug-07 & -- &   Mar-08      &&     \multirow{2}{*}{Oct-07} & \multirow{2}{*}{--} & \multirow{2}{*}{Feb-10} && \multirow{2}{*}{Aug-07} & \multirow{2}{*}{--} & \multirow{2}{*}{Nov-09} \\      
May-08 & -- &   Feb-10      &&      \\
May-10 & -- &   Jul-10      &&     Jun-10 & -- &   Mar-11     &&  \multirow{3}{*}{Apr-10} & \multirow{3}{*}{--} & \multirow{3}{*}{Dec-12} \\
Aug-11 & -- &   Aug-12      &&      Aug-11 & -- &   Aug-12      &&      \\
Nov-12 & -- &   Dec-12      &&      Nov-12 & -- &   Jan-13      &&      \\
 \bottomrule
\end{tabularx}
\label{Summary of Critical Periods}
\end{table} 

In Figure \ref{Network Density}, we present a plot of the network densities of the competing methods over a yearly sample moving window. 
In general, the BCGLPM(1) process recorded a more dense structure, followed by the BCGLPM(0) process, with the SSSL being more sparse. 
Following the concept of shape variation analysis \cite[see][]{Dryden_2016}, we take closer look at the properties of the network densities invariant to scale, translation and rotation. 
The commonest approach to achieve this goal is through standardization (or normalization).  
Figure \ref{Standardized Network Density} presents the standardized network densities over the yearly sample moving window. 

When thresholding the network density for contagion analysis, the level of inferred vulnerability tend to be highly sensitive to the choice of threshold. However, using 
historical events and crises time lines, we define the dashed zero-line as a natural choice of a threshold where negative densities suggest bearable connectedness
and positive densities indicate potential systemic vulnerability for shock amplification. Based on this definition, 
we present in Table \ref{Summary of Critical Periods}, the summary of the critical sample periods with positive standardized network densities. 
From the figure and the table, we observe that one major difference between the densities of the three schemes is that SSSL and the BCGLPM(0) crisscrossed the zero-line line many times 
compared to that of the BCGLPM(1).  

In analyzing the shape of the standardized densities, a key difference in the three plots emerges in the period preceding the financial crisis. The SSSL and the 
BCGLPM(0) recorded relatively low densities during the 2004-2007 period.    
Many empirical financial network papers that use stock market data have findings similar to that of the SSSL and BCGLPM(0). Based on these low densities, 
many researchers refer to the 2004-2007 sub-period as a time of calm market conditions. However, the result of the BCGLPM(1) seems to differ from this assertion by reporting 
a steady rise from mid-2005 to mid-2007.  

Now, according to the report by the \cite{Financial_2011} (FCIC), in the years preceding the crisis, the US experienced low-interest rates and inflow of money. This  
allowed banks to lower rates and requirements, giving way to easy lending even to consumers with less or no credit to acquire risky mortgages in the anticipation that they would be able to quickly 
refinance at easier terms. The sub-prime mortgage origination over the sub-period of 2003 to 2006 increased from a low of 8.3\% to an incredible level of 23.5\%. This and other existing conditions contributed to 
many financial firms holding highly related (mortgage) securities which led to a higher correlation among many institutions. 
The report also revealed that some firms publicly issued signals that reassured the market (rating agencies, investors, and clients) while withholding certain information about 
firm specific deals. Thus, firms had a private view of the market which differed from the market view of firms. It is therefore not surprising that 2004-2007 appears calm from the market perspective, 
as reported by the SSSL. Tracking the firm-level information, however, reveals a development in the density reported by the BCGLPM(1) from mid-2005 to mid-2007 before taking an upward turn.   

During late-2007 to late-2009, we observe that the density of all three methods where above the zero-line. Relating this to events in the financial system, we notice that around 2007, the many US financial institutions experienced a
liquidity crisis following the fall in housing prices and an abrupt shutdown of sub-prime lending. This led to losses for many financial institutions who held mortgage-related securities. Such events disrupted 
several financial market operations creating a cascade of the sale of securities by many institutions, which lowered their values and increased their volatility connectedness.  
In 2008, the financial system was thrown into turmoil with the near collapse and acquisition of Bear Sterns by JP Morgan Chase in March 2008, followed by the bankruptcy of Lehman Brothers 
(the fourth largest U.S. investment bank at the time) and the bailout of American International Group (AIG - the world's largest insurance company) in September 2008. These events 
triggered other actions of market participants affecting a broader aspect of the US system and many economies and institutions across the globe resulting in a global recession. 

Between late-2009 to late-2011, we notice another difference in the standardized densities. Here, the SSSL reported values below the zero-line whiles the BCGLPM(1) 
shows a steady increase in volatility connectedness. According to the time line of events in the financial system, 2010--2012 sub-period coincides with a time of struggle among Euro area members 
to recover from the global recession. Unlike the US, the recovery process for the Euro zone was much more difficult due to its organizational structure of having a common monetary policy but different fiscal measures. 
Furthermore, the EU was thrown into a crisis that centered on heavily indebted countries like Greece, Ireland, Portugal, Spain, and Italy. 
This became tense in early-2010 when Greece government bonds were downgraded and 
a bailout request by Greece was met with a plan by the EU and the IMF that was supposed to take effect over the following three years. In late-2010, Greece, Ireland, and Portugal were 
reported to have high fiscal related problems. These events sent a strong signal to investors and financial institutions that held 
European sovereign bonds. In mid-2011, the threat to European financial institutions and the global financial system became severe when the crisis of Greece, Ireland, and Portugal began to affect Spain and Italy 
(the third largest Euro zone economy and second biggest debtor to bond investors). More importantly, many European institutions were heavily exposed to Spain and Italy, thus, spreading the crisis within and 
beyond Europe. 

From the figure, we see that after 2009, the next peak of all the standardized densities was recorded around early-2012. According to unfolding events in the global financial system, early-2012 coincided  
with S\&P downgrading nine Euro area countries including Austria, Cyprus, France, Italy, Malta, Portugal, Slovakia, Slovenia and Spain. With France being the second Euro zone economy, this exercise had a strong 
effect on many investors, financial institutions and markets leading to an increased volatility connectedness.

From the discussion so far, we notice that by tracking the idiosyncratic components of stock volatilities of firms, the BCGLPM(1) is able to identify and relate the vulnerabilities that propagate the 
triggers of the crisis to actions of financial institutions that are not directly observed by the market. Furthermore, there is a lagged market response to these actions which only become apparent 
during the crisis.

\subsection{Procrustean Analysis of Node Positions}

In order to gain insight into the information inherent in the inferred latent nodal positions, we compare the configurations of these positions via Procrustean analysis. 
This approach is based on a comparison of two coordinate matrices (with corresponding points), with the goal of transforming the second set of coordinates to be as close as possible to the first. 
The transformation involves a combination of translating, uniform scaling, rotating and reflection. The goodness-of-fit is measured by the residual sum of squares, referred to as the Procrustes distance statistics.  
The Procrustes approach has been applied in many fields for investigating similarity of sets of spatial positions \cite[see][]{Dryden_2016, Cox_2000, Gower_2004, Wang_2010}.  
It is useful here in allowing us to compare latent positions that are unique only up to rotation.

Let $U^0$ and $U^1$ denote the coordinates of the node positions estimated by the BCGLPM(0) and BCGLPM(1) respectively. We compute $\hat{U}$ as the Procrustean transformation of $U^1$ with $U^0$ as the 
target. The transformation is generally expressed as:
\begin{align}
 \hat{U} = \rho U^1 H + c
\end{align}
where $\rho$ is a scalar dilation, $H$ is a $2\times 2$ orthogonal matrix representing a rotation and reflection, and $c$ is a $2\times 1$ translation vector. 
The Procrustean statistic, $D(U^0, U^1)$, is obtained as the standardized distance between the target, $U^0$ and the transformed $U^1$, that is $\hat{U}$. This statistic 
takes a minimum of $0$ and a maximum of $1$, where values close to zero (one) indicate (dis)similarity in the sets of latent positions. 

\begin{figure}[!ht]
\footnotesize
\centering
\includegraphics[height=0.22\textheight]{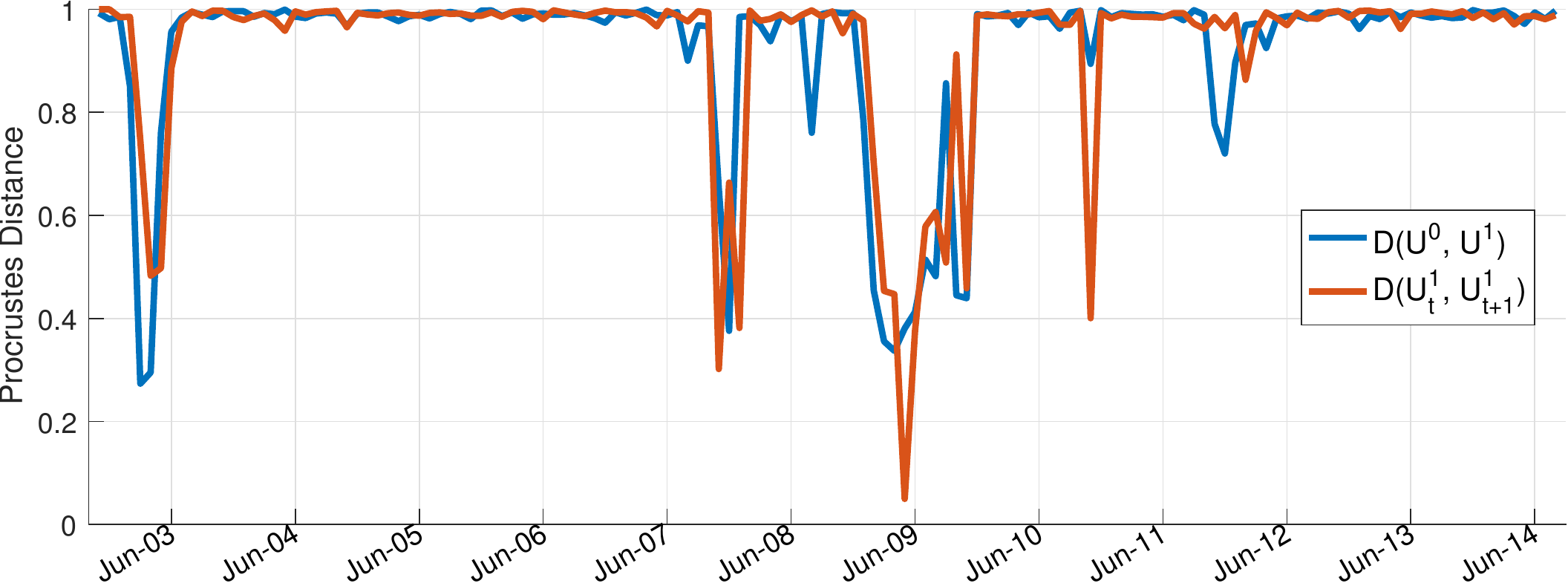}  
\caption{A Procrustes distance plot of $D(U^0, U^1)$ in blue and $D(U^1_t, U^1_{t+1})$ in red. The $D(U^0, U^1)$ plot measures the dissimilarity between 
 $U^0$ and $U^1$ estimated by the BCGLPM(0) and BCGLPM(1) respectively, and $D(U^1_t, U^1_{t+1})$ plot measures the change in positions, $U^1_t$ and $U^1_{t+1}$ of BCGLPM(1) 
 over the sample period. Note: The smaller the distance (close to zero) the similar the nodal positions.}
\label{Procrustean Distance}
\end{figure}

Figure \ref{Procrustean Distance} presents a plot of the Procrustean distance $D(U^0, U^1)$ and $D(U^1_t, U^1_{t+1})$. The plot of $D(U^0, U^1)$ measures the dissimilarity between 
 $U^0$ and $U^1$ estimated by the BCGLPM(0) and BCGLPM(1) respectively, and $D(U^1_t, U^1_{t+1})$ measures the change in positions, $U^1_t$ and $U^1_{t+1}$ of BCGLPM(1). 
The change in $U^1_t$ and $U^1_{t+1}$ appears to be closely related to that of the comparison the dissimilarity between $U^0$ and $U^1$. Both plots display 
highly dissimilar positions of firms over greater part of the sample period. However, we notice some exceptions. By comparing Figure \ref{Procrustean Distance} and the standardized density plot 
in Figure \ref{Standardized Network Density}, we see that the portion of times of high changes in the pairs $(U^0, U^1)$ and $(U^1_t, U^1_{t+1})$ coincides with periods of negative densities, 
whiles the times of relatively stable positions appears to occur in periods of positive densities, e.g., early to mid-2003, late-2006 to early 2007, late 2007,  2008-2009, early 2010, and late 2010 to early 2011. 
We can therefore infer from these observation that, firms positions in financial markets changes frequently when the system-wide connectedness is at bearable levels, and remains relatively 
stable in critical times of higher systemic vulnerabilities. 

In normal times, financial firms tend to take different positions when buying or selling stocks, bonds, currencies, commodities or other financial instruments traded by other 
institutions, investors or clients. Such positions are 
often necessary to make profits or obtain benefits from the movements in the market. However, in critical periods when firms perceives that crisis is lurking and the market for certain commodities may soon be abandoned, 
or that doing business with some institutions would be too risky due to their exposures to some risk, such firms tend to take entrenched positions by cutting off ties with other partners or from 
trading particular commodities. Some firms proceed further by sending signals to many others and regulators about such potential trouble institutions or commodities. 
As reported by the FCIC, in 2007, the problems of the US financial 
market hit the largest French bank, BNP Paribas and many others. In response, BNP sent a strong shivering signal to other big market players and regulators about the US securities market. 
This led to disruptions in the securities market which quickly spread to other parts of the money market. A a result, the market stopped trading certain commodities, lenders quickly withdrew from many programs, and 
investors dumped their security holdings and increased their holdings in seemingly safer money market funds and treasury bonds.

\subsection{Dynamics of Firm Clustering Behaviors}
\begin{figure}[!ht]
\footnotesize
\centering
\includegraphics[height=0.25\textheight]{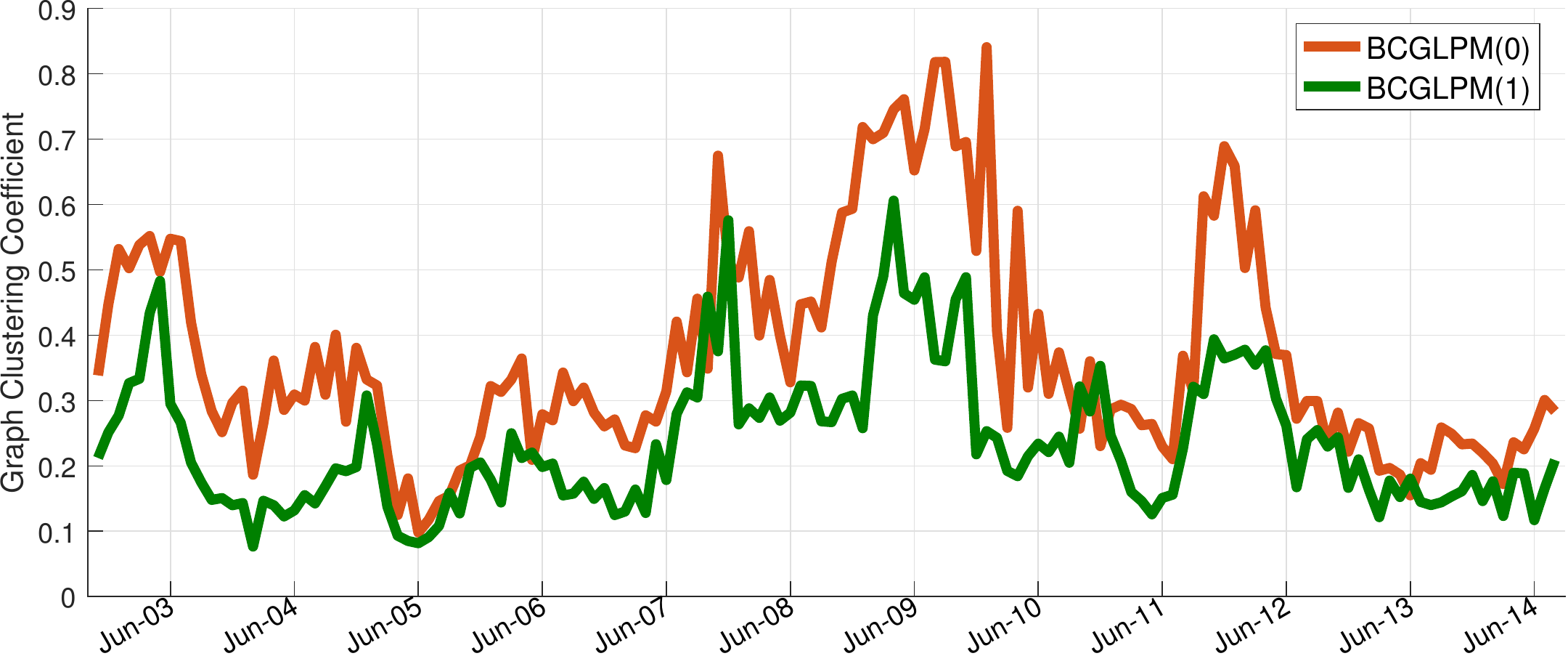}  
\caption{Dynamics of the Graph Clustering Coefficient (GCC) of the BCGLPM(0) in red and BCGLPM(1) in green, obtained from a one-year moving window estimation 
 from January, 2002 - August, 2014.}
\label{Clustering Plot}
\end{figure}

We study the clustering behavior of the firms by monitoring the dynamics of the network clustering coefficient according to~\citep[see][]{Barrat_2000} and its relevance to contagion in financial networks.  
Figure \ref{Clustering Plot} presents the graph clustering coefficient. This notion of clustering corresponds to the social network concept of transitivity coefficient and can be captured
numerically as follows:
\begin{align}
 GCC = \frac{3 \times \text{(number of triangles)} }{ \text{(number of open triads)}}
\end{align}
where open triads is defined as a connected subgraph consisting of three nodes and two edges. 
$GCC$ takes values between 0 and 1, and measures the tendency for nodes in a network form clusters or triangles.
We observe from Figure \ref{Clustering Plot} that the transitivity coefficient is quite high in turbulent times, with the $GCC$ index significantly higher than 0.3 in almost all cases.

\begin{figure}[!ht]
\centering
\subfloat[ ]{
\includegraphics[height=0.25 \textheight]{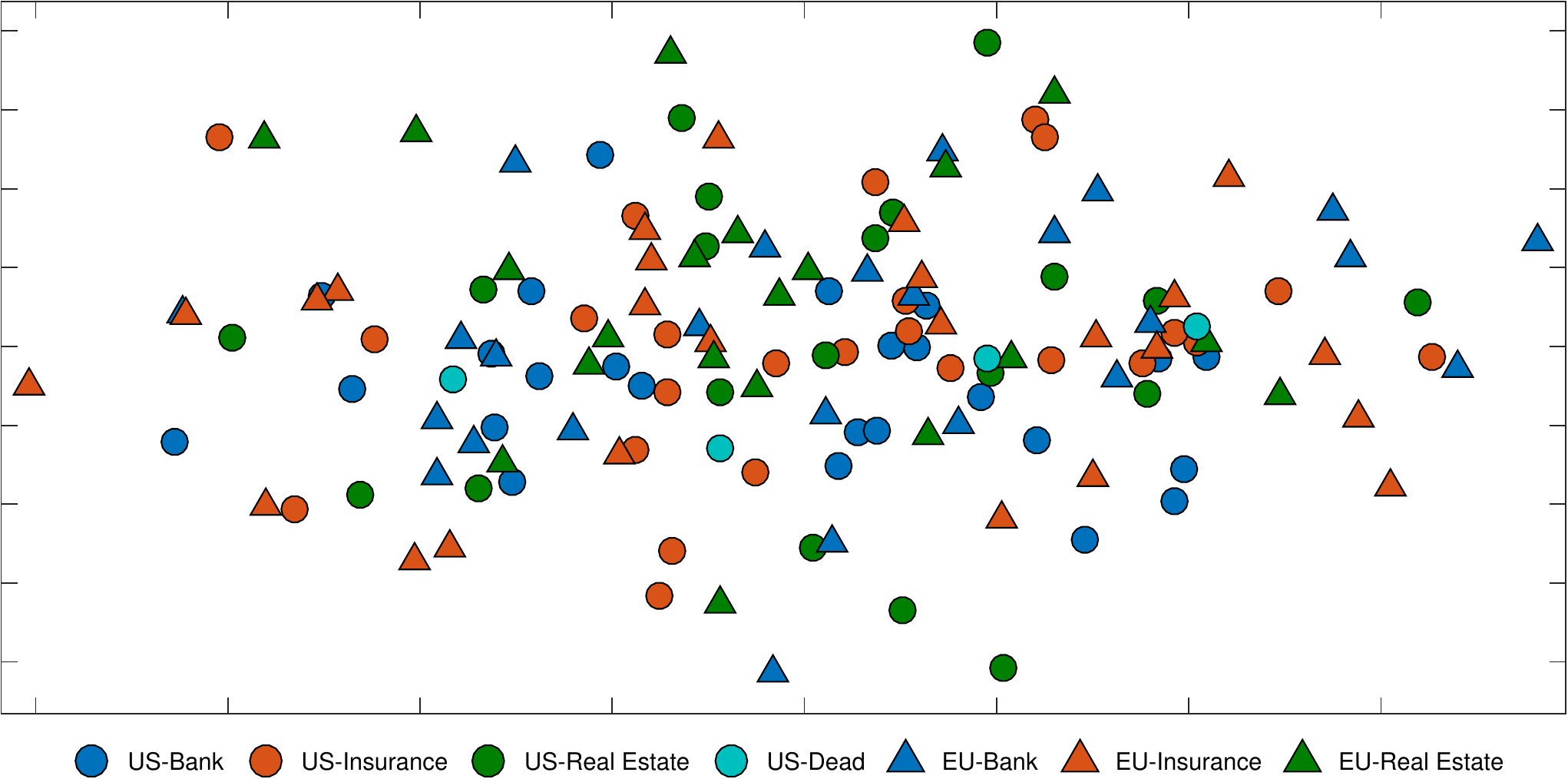}   
\label{Win 1}}  \qquad 
\subfloat[ ]{
\includegraphics[height=0.25 \textheight]{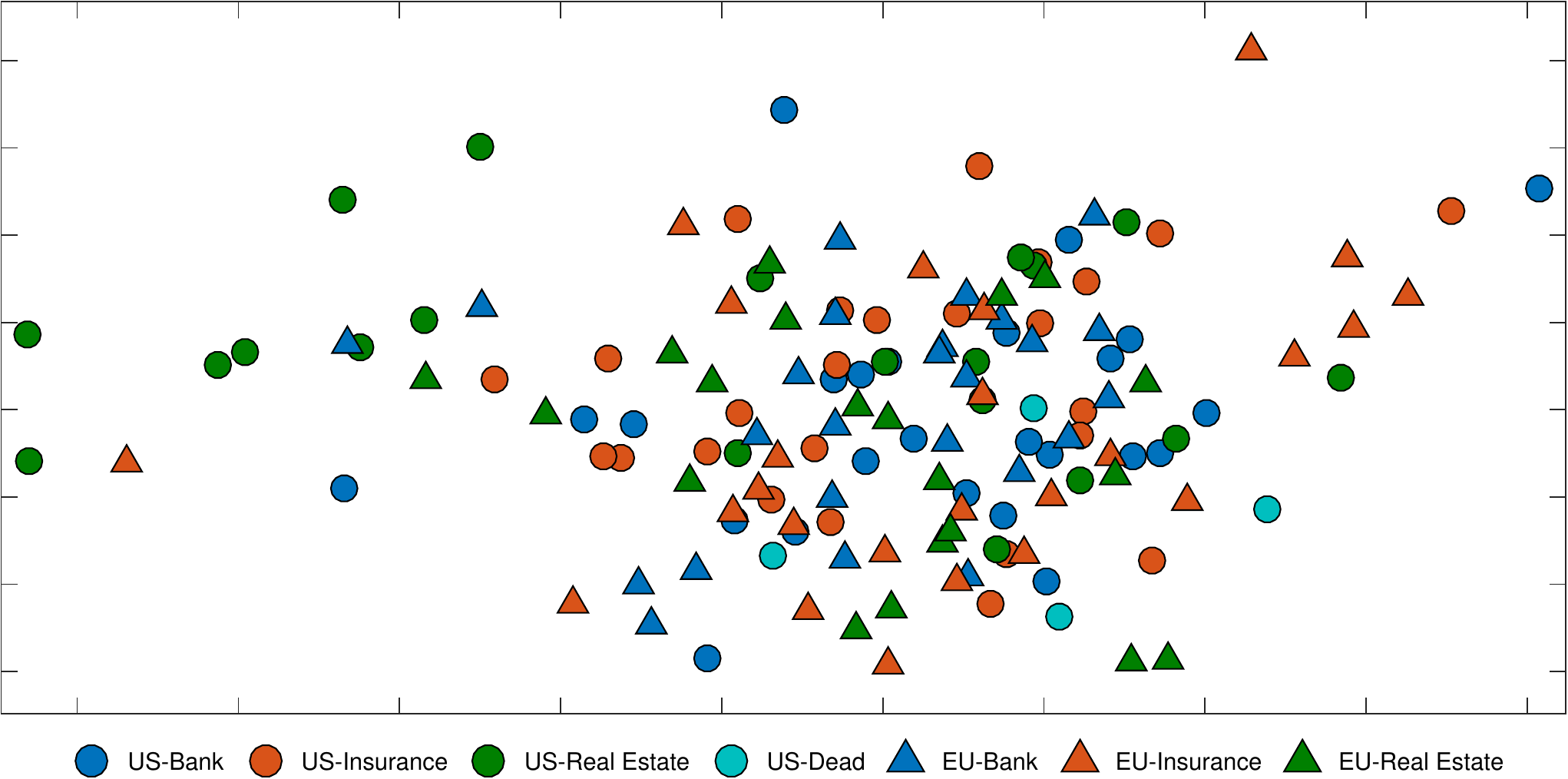}     
\label{Win 2}} \qquad 
\subfloat[ ]{
\includegraphics[height=0.25 \textheight]{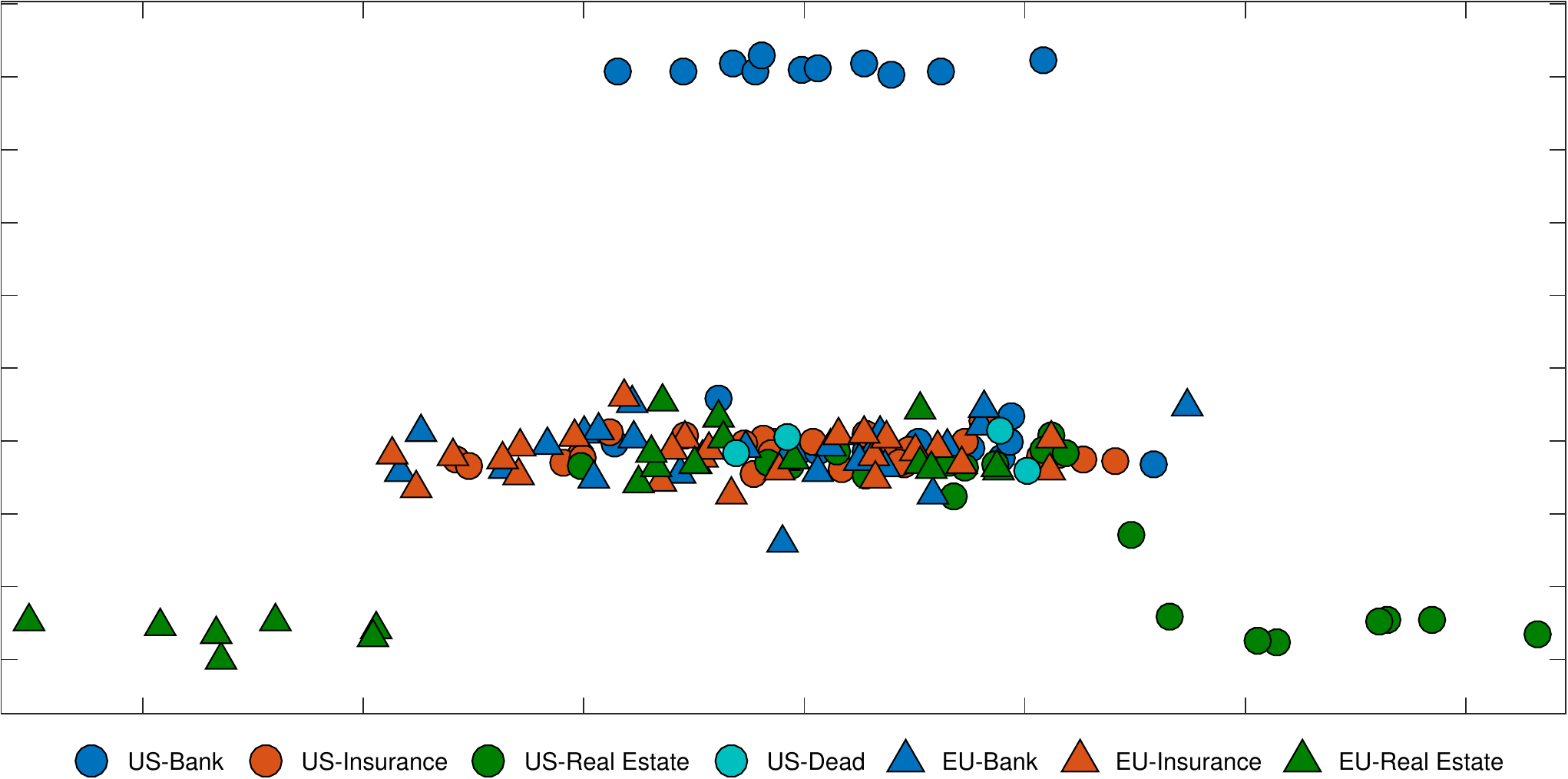} 
\label{Win 3}} 
\caption{A spatial representation of the latent positions for the moving window ending (\ref{Win 1}) June 2007 ($GCC$ = 0.18), 
(\ref{Win 2}) August 2007 ($GCC$ = 0.31), and  (\ref{Win 3}) December 2007 ($GCC$ = 0.58).}
\label{SUB_1}
\end{figure}

\begin{figure}[!ht]
\centering
\subfloat[ ]{
\includegraphics[height=0.25 \textheight]{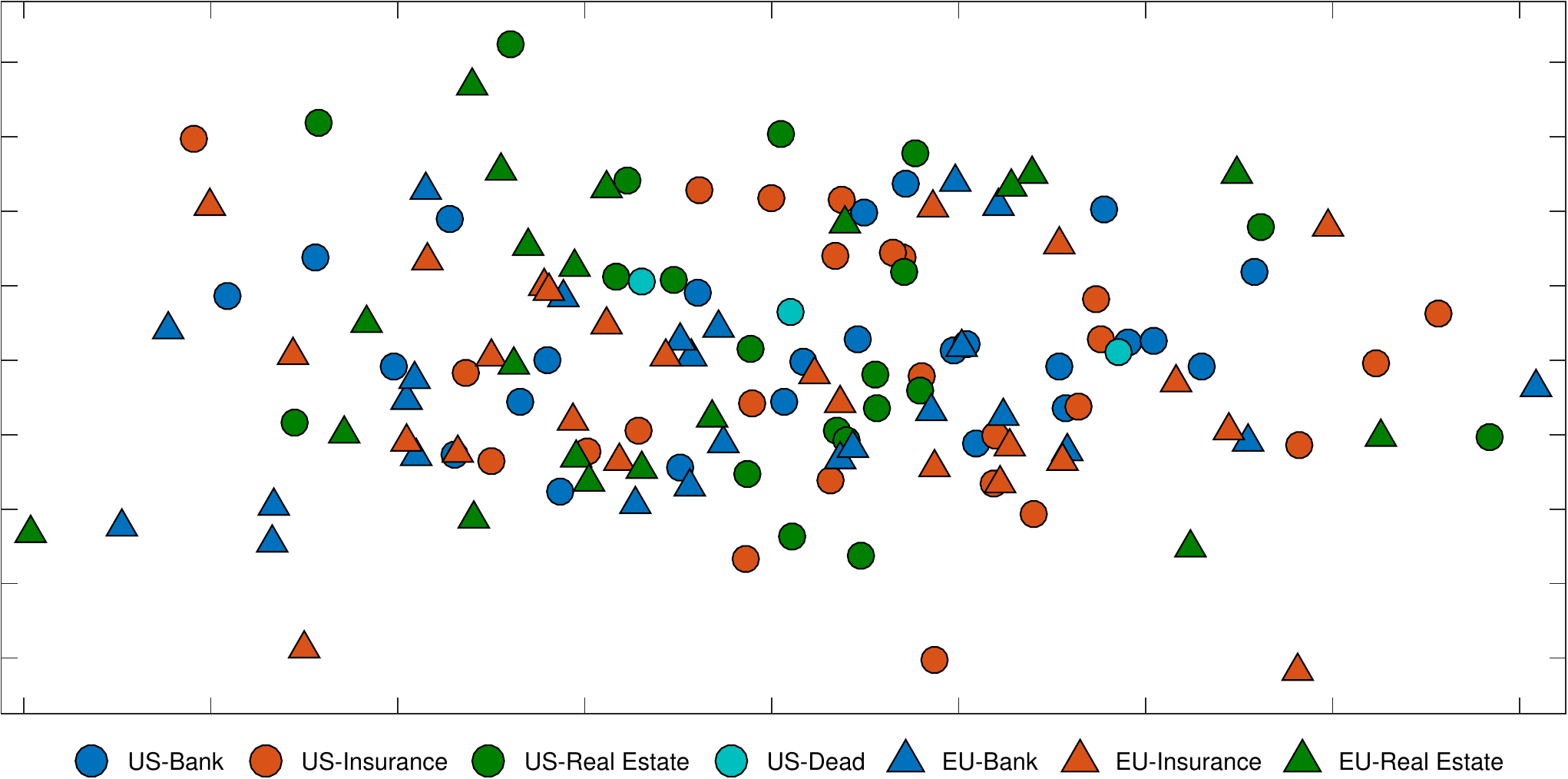} 
\label{Win 4}} \qquad
\subfloat[ ]{
\includegraphics[height=0.25 \textheight]{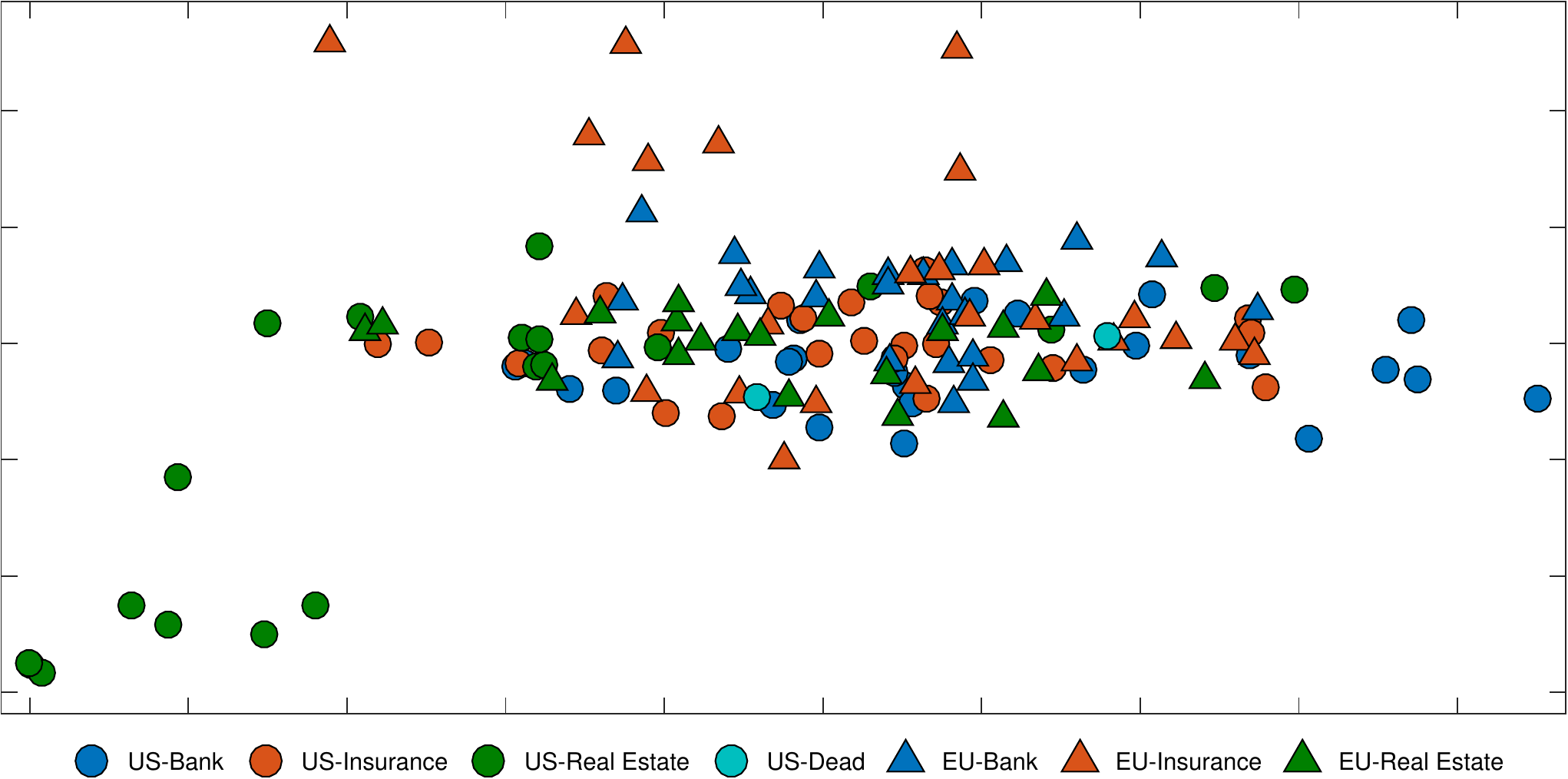}     
\label{Win 5}}  \qquad
\subfloat[ ]{
\includegraphics[height=0.25 \textheight]{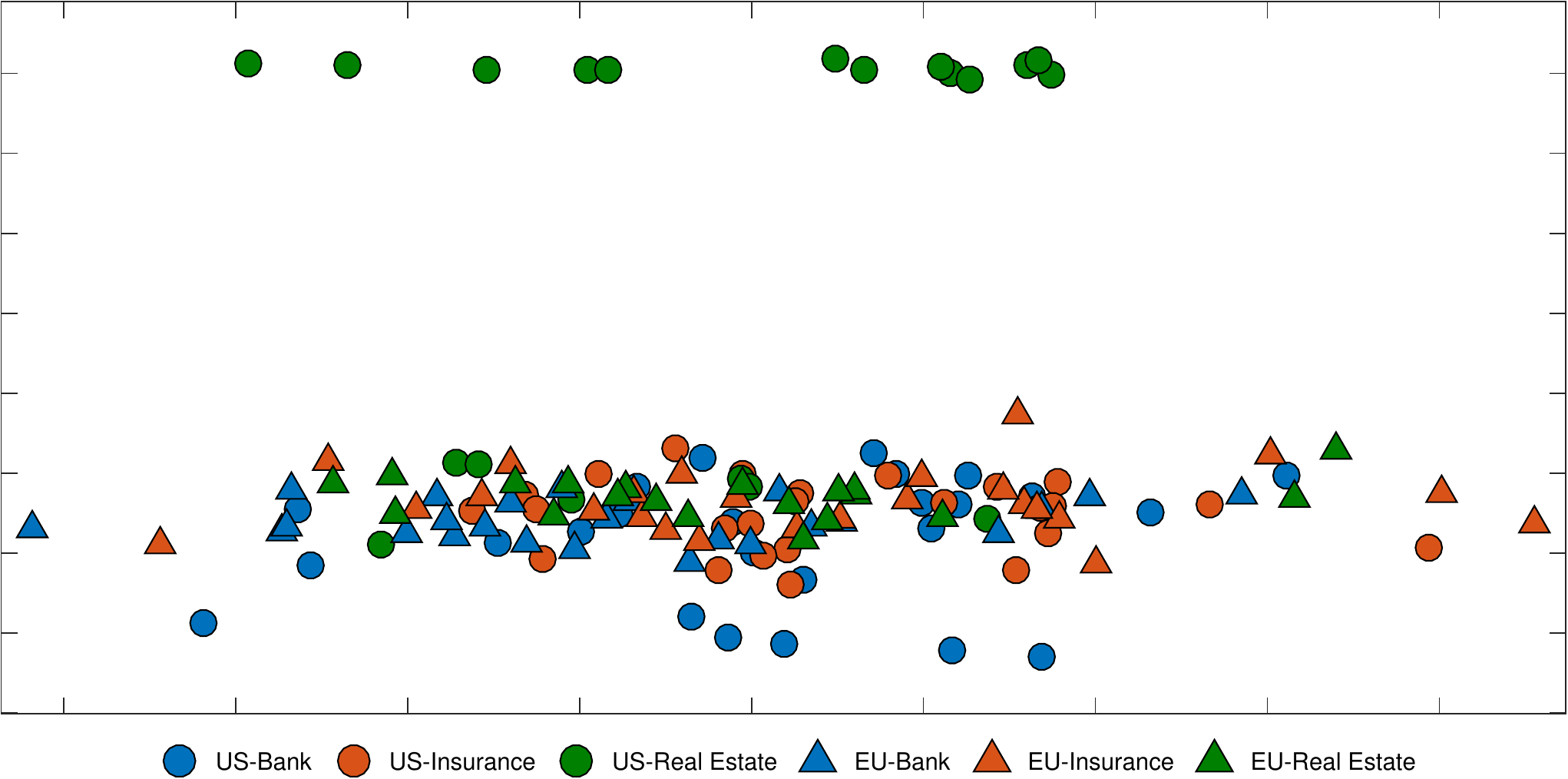} 
\label{Win 6}}
\caption{A spatial representation of the latent positions for the moving window ending (\ref{Win 4}) June 2008 ($GCC$ = 0.28), 
(\ref{Win 5}) August 2008 ($GCC$ = 0.32), and  (\ref{Win 6}) April 2009 ($GCC$ = 0.61).}
\label{SUB_2}
\end{figure}

To relate the structural trends to firm clustering behaviors during the 2007-2009 sub-period, we present in Figures \ref{SUB_1} and \ref{SUB_2}, 
the spatial position of the latent features of the institutions from mid-2007 to early-2009. 
US institutions are represented with circle nodes whiles European institutions are in triangle nodes. The colors represent the industries in the sense that blue denote banks, red for insurance, 
green for real estates and cyan for acquired or bankrupted institutions. 

We notice a significant change in the transitivity coefficient between June 2007 through August 2007 to December 2007, and similarly between June 2008 through August 2008 to April 2009. 
In the structure for December 2007 (see Figure \ref{Win 3}), we notice a cluster of US Banks (blue circles) at the top, EU Real Estates (green triangles) on the south west corner and  
US Real Estates (green circles) on the south east corner with the rest concentrated in the center. Likewise, the structure for April 2009 (see Figure \ref{Win 6}) also depicts a cluster of 
US Real Estates (green circles) at the top with the rest concentrated in the center. These two periods are also associated with higher transitivity index exceeding 0.5. 
This shows that periods of high transitivity in the network structure coincided with clustering behaviors among firms in specific industries based on 
the spatial representation of the nodes in the network.
\section{Conclusion}
\label{Conclusion}

We consider a Bayesian covariance graphical and latent space approach to model financial linkages where nodes representing financial institutions are uniformly distributed over a two-dimensional plane. 
The approach is based on modeling the idiosyncratic channel of exposures among firms from a multivariate financial time series via covariance graphical models (CGMs) and uncovering the nodal positions via latent position
models (LPMs). We present an efficient Markov chain Monte Carlo algorithm to iterate between sampling parameters of the CGM and the LPM, using samples from the latter to
update prior information for covariance graph selection.

We study the dynamics of the exposure among 150 publicly traded top financial institutions in the United States and Europe between January 2002 to August 2014. The goal is to 
uncover how the latent spatial positions contribute to the clustering behavior of firms and the spread of financial risk. 
By tracking the idiosyncratic volatility linkages, we find evidence of a rising volatility connectedness from mid-2005 to mid-2007 before taking an upward turn after August 2007. 
In monitoring the latent position of firms, we found evidence that in normal times, financial firms tend to take different positions when trading financial instruments. 
However, in periods preceding financial turbulence, many of these firms tend to take entrenched positions to cut ties with those they consider too risky to avoid exposures.

\bibliographystyle{imsart-nameyear}
\bibliography{Bibliography.bib}

\begin{thebibliography}{32}

\bibitem[\protect\citeauthoryear{Acemoglu, Ozdaglar and
  Tahbaz-Salehi}{2015}]{Acemoglu_2015}
\begin{barticle}[author]
\bauthor{\bsnm{Acemoglu},~\bfnm{Daron}\binits{D.}},
  \bauthor{\bsnm{Ozdaglar},~\bfnm{Asuman}\binits{A.}} \AND
  \bauthor{\bsnm{Tahbaz-Salehi},~\bfnm{Alireza}\binits{A.}}
(\byear{2015}).
\btitle{{Systemic Risk and Stability in Financial Networks}}.
\bjournal{American Economic Review}
\bvolume{105}
\bpages{564--608}.
\end{barticle}
\endbibitem

\bibitem[\protect\citeauthoryear{Ahelegbey, Billio and
  Casarin}{2016}]{Ahelegbey_2016}
\begin{barticle}[author]
\bauthor{\bsnm{Ahelegbey},~\bfnm{Daniel~Felix}\binits{D.~F.}},
  \bauthor{\bsnm{Billio},~\bfnm{Monica}\binits{M.}} \AND
  \bauthor{\bsnm{Casarin},~\bfnm{Roberto}\binits{R.}}
(\byear{2016}).
\btitle{{Bayesian Graphical Models for Structural Vector Autoregressive
  Processes}}.
\bjournal{Journal of Applied Econometrics}
\bvolume{31}
\bpages{357--386}.
\end{barticle}
\endbibitem

\bibitem[\protect\citeauthoryear{Arregui et~al.}{2013}]{Arregui_2013}
\begin{bbook}[author]
\bauthor{\bsnm{Arregui},~\bfnm{Nicolas}\binits{N.}},
  \bauthor{\bsnm{Norat},~\bfnm{Mohamed}\binits{M.}},
  \bauthor{\bsnm{Pancorbo},~\bfnm{Antonio}\binits{A.}},
  \bauthor{\bsnm{Scarlata},~\bfnm{Jodi~G}\binits{J.~G.}},
  \bauthor{\bsnm{Holttinen},~\bfnm{Eija}\binits{E.}},
  \bauthor{\bsnm{Melo},~\bfnm{Fabiana}\binits{F.}},
  \bauthor{\bsnm{Surti},~\bfnm{Jay}\binits{J.}},
  \bauthor{\bsnm{Wilson},~\bfnm{Christopher}\binits{C.}},
  \bauthor{\bsnm{Wehrhahn},~\bfnm{Rodolfo}\binits{R.}} \AND
  \bauthor{\bsnm{Yanase},~\bfnm{Mamoru}\binits{M.}}
(\byear{2013}).
\btitle{{Addressing Interconnectedness: Concepts and Prudential Tools}}.
\bpublisher{International Monetary Fund}.
\end{bbook}
\endbibitem

\bibitem[\protect\citeauthoryear{Barrat and Weigt}{2000}]{Barrat_2000}
\begin{barticle}[author]
\bauthor{\bsnm{Barrat},~\bfnm{Alain}\binits{A.}} \AND
  \bauthor{\bsnm{Weigt},~\bfnm{Martin}\binits{M.}}
(\byear{2000}).
\btitle{{On The Properties of Small-World Network Models}}.
\bjournal{The European Physical Journal B-Condensed Matter and Complex Systems}
\bvolume{13}
\bpages{547--560}.
\end{barticle}
\endbibitem

\bibitem[\protect\citeauthoryear{Bernanke}{2013}]{Bernanke_2013}
\begin{btechreport}[author]
\bauthor{\bsnm{Bernanke},~\bfnm{Ben}\binits{B.}}
(\byear{2013}).
\btitle{{Monitoring the Financial System}}
\btype{Speech},
\bpublisher{at the 49th Annual Conference Bank Structure and Competition
  sponsored by the Federal Reserve Bank of Chicago, May 10}.
\end{btechreport}
\endbibitem

\bibitem[\protect\citeauthoryear{Billio et~al.}{2012}]{Billio_2012}
\begin{barticle}[author]
\bauthor{\bsnm{Billio},~\bfnm{M}\binits{M.}},
  \bauthor{\bsnm{Getmansky},~\bfnm{M}\binits{M.}},
  \bauthor{\bsnm{Lo},~\bfnm{A.~W}\binits{A.~W.}} \AND
  \bauthor{\bsnm{Pelizzon},~\bfnm{L}\binits{L.}}
(\byear{2012}).
\btitle{{Econometric Measures of Connectedness and Systemic Risk in the Finance
  and Insurance Sectors}}.
\bjournal{Journal of Financial Economics}
\bvolume{104}
\bpages{535 -- 559}.
\end{barticle}
\endbibitem

\bibitem[\protect\citeauthoryear{Brubaker, Salzmann and
  Urtasun}{2012}]{Brubaker_2012}
\begin{binproceedings}[author]
\bauthor{\bsnm{Brubaker},~\bfnm{Marcus~A}\binits{M.~A.}},
  \bauthor{\bsnm{Salzmann},~\bfnm{Mathieu}\binits{M.}} \AND
  \bauthor{\bsnm{Urtasun},~\bfnm{Raquel}\binits{R.}}
(\byear{2012}).
\btitle{{A Family of MCMC Methods on Implicitly Defined Manifolds}}.
In \bbooktitle{AISTATS}
\bpages{161--172}.
\end{binproceedings}
\endbibitem

\bibitem[\protect\citeauthoryear{{Financial Crisis Inquiry
  Commission}}{2011}]{Financial_2011}
\begin{bbook}[author]
\bauthor{\bsnm{{Financial Crisis Inquiry Commission}}}
(\byear{2011}).
\btitle{{The Financial Crisis Inquiry Report: The Final Report of the National
  Commission on the Causes of the Financial and Economic Crisis in the United
  States}}.
\bpublisher{Public Affairs}.
\end{bbook}
\endbibitem

\bibitem[\protect\citeauthoryear{Cox and Cox}{2000}]{Cox_2000}
\begin{bbook}[author]
\bauthor{\bsnm{Cox},~\bfnm{Trevor~F}\binits{T.~F.}} \AND
  \bauthor{\bsnm{Cox},~\bfnm{Michael~AA}\binits{M.~A.}}
(\byear{2000}).
\btitle{{Multidimensional Scaling}}.
\bpublisher{CRC press}.
\end{bbook}
\endbibitem

\bibitem[\protect\citeauthoryear{Diebold and Yilmaz}{2014}]{Diebold_2014}
\begin{barticle}[author]
\bauthor{\bsnm{Diebold},~\bfnm{F.}\binits{F.}} \AND
  \bauthor{\bsnm{Yilmaz},~\bfnm{K}\binits{K.}}
(\byear{2014}).
\btitle{{On the Network Topology of Variance Decompositions: Measuring the
  Connectedness of Financial Firms}}.
\bjournal{Journal of Econometrics}
\bvolume{182}
\bpages{119--134}.
\end{barticle}
\endbibitem

\bibitem[\protect\citeauthoryear{Dryden and Mardia}{2016}]{Dryden_2016}
\begin{bbook}[author]
\bauthor{\bsnm{Dryden},~\bfnm{Ian~L}\binits{I.~L.}} \AND
  \bauthor{\bsnm{Mardia},~\bfnm{Kanti~V}\binits{K.~V.}}
(\byear{2016}).
\btitle{{Statistical Shape Analysis: With Applications in R}}.
\bpublisher{John Wiley \& Sons}.
\end{bbook}
\endbibitem

\bibitem[\protect\citeauthoryear{Dungey and Gajurel}{2015}]{Dungey_2015}
\begin{barticle}[author]
\bauthor{\bsnm{Dungey},~\bfnm{Mardi}\binits{M.}} \AND
  \bauthor{\bsnm{Gajurel},~\bfnm{Dinesh}\binits{D.}}
(\byear{2015}).
\btitle{{Contagion and Banking Crisis--International Evidence for 2007--2009}}.
\bjournal{Journal of Banking and Finance}
\bvolume{60}
\bpages{271--283}.
\end{barticle}
\endbibitem

\bibitem[\protect\citeauthoryear{Elliott, Golub and
  Jackson}{2014}]{Elliott_2014}
\begin{barticle}[author]
\bauthor{\bsnm{Elliott},~\bfnm{Matthew}\binits{M.}},
  \bauthor{\bsnm{Golub},~\bfnm{Benjamin}\binits{B.}} \AND
  \bauthor{\bsnm{Jackson},~\bfnm{Matthew~O}\binits{M.~O.}}
(\byear{2014}).
\btitle{{Financial Networks and Contagion}}.
\bjournal{American Economic Review}
\bvolume{104}
\bpages{3115--3153}.
\end{barticle}
\endbibitem

\bibitem[\protect\citeauthoryear{Gelman and Rubin}{1992}]{Gelman_1992}
\begin{barticle}[author]
\bauthor{\bsnm{Gelman},~\bfnm{A.}\binits{A.}} \AND
  \bauthor{\bsnm{Rubin},~\bfnm{D.~B.}\binits{D.~B.}}
(\byear{1992}).
\btitle{{Inference from Iterative Simulation Using Multiple Sequences, (with
  discussion)}}.
\bjournal{Statistical Science}
\bvolume{7}
\bpages{457--511}.
\end{barticle}
\endbibitem

\bibitem[\protect\citeauthoryear{George, Sun and Ni}{2008}]{George_2008}
\begin{barticle}[author]
\bauthor{\bsnm{George},~\bfnm{Edward~I.}\binits{E.~I.}},
  \bauthor{\bsnm{Sun},~\bfnm{Dongchu}\binits{D.}} \AND
  \bauthor{\bsnm{Ni},~\bfnm{Shawn}\binits{S.}}
(\byear{2008}).
\btitle{{Bayesian Stochastic Search for VAR Model Restrictions}}.
\bjournal{Journal of Econometrics}
\bvolume{142}
\bpages{553--580}.
\end{barticle}
\endbibitem

\bibitem[\protect\citeauthoryear{Gower and Dijksterhuis}{2004}]{Gower_2004}
\begin{bbook}[author]
\bauthor{\bsnm{Gower},~\bfnm{John~C}\binits{J.~C.}} \AND
  \bauthor{\bsnm{Dijksterhuis},~\bfnm{Garmt~B}\binits{G.~B.}}
(\byear{2004}).
\btitle{{Procrustes Problems}}
\bvolume{30}.
\bpublisher{Oxford University Press on Demand}.
\end{bbook}
\endbibitem

\bibitem[\protect\citeauthoryear{Hoff}{2008}]{Hoff_2008}
\begin{binproceedings}[author]
\bauthor{\bsnm{Hoff},~\bfnm{Peter~D}\binits{P.~D.}}
(\byear{2008}).
\btitle{{Modeling Homophily and Stochastic Equivalence in Symmetric Relational
  Data}}.
In \bbooktitle{Advances in Neural Information Processing Systems}
\bpages{657--664}.
\end{binproceedings}
\endbibitem

\bibitem[\protect\citeauthoryear{Hoff}{2009}]{Hoff_2009}
\begin{barticle}[author]
\bauthor{\bsnm{Hoff},~\bfnm{Peter~D}\binits{P.~D.}}
(\byear{2009}).
\btitle{{Simulation of the Matrix Bingham-von Mises-Fisher Distribution, with
  Applications to Multivariate and Relational Data}}.
\bjournal{Journal of Computational and Graphical Statistics}
\bvolume{18}
\bpages{438--456}.
\end{barticle}
\endbibitem

\bibitem[\protect\citeauthoryear{Hoff}{2013}]{Hoff_2013}
\begin{bunpublished}[author]
\bauthor{\bsnm{Hoff},~\bfnm{Peter~D}\binits{P.~D.}}
(\byear{2013}).
\btitle{{Bayesian Analysis of Matrix Data with Rstiefel}}.
\bnote{R-vignette}.
\end{bunpublished}
\endbibitem

\bibitem[\protect\citeauthoryear{IMF}{2011}]{IMF_2011}
\begin{btechreport}[author]
\bauthor{\bsnm{IMF}}
(\byear{2011}).
\btitle{{Global Financial Stability Report: Grappling with Crisis Legacies}}
\btype{Technical Report},
\bpublisher{World Economic and Financial Services}.
\end{btechreport}
\endbibitem

\bibitem[\protect\citeauthoryear{Kolaczyk and Cs{\'a}rdi}{2014}]{Kolaczyk_2014}
\begin{bbook}[author]
\bauthor{\bsnm{Kolaczyk},~\bfnm{Eric~D}\binits{E.~D.}} \AND
  \bauthor{\bsnm{Cs{\'a}rdi},~\bfnm{G{\'a}bor}\binits{G.}}
(\byear{2014}).
\btitle{{Statistical Analysis of Network Data with R}}
\bvolume{65}.
\bpublisher{Springer}.
\end{bbook}
\endbibitem

\bibitem[\protect\citeauthoryear{Lenkoski and Dobra}{2011}]{Lenkoski_2011}
\begin{barticle}[author]
\bauthor{\bsnm{Lenkoski},~\bfnm{Alex}\binits{A.}} \AND
  \bauthor{\bsnm{Dobra},~\bfnm{Adrian}\binits{A.}}
(\byear{2011}).
\btitle{{Computational Aspects Related to Inference in Gaussian Graphical
  Models With the G-Wishart Prior}}.
\bjournal{Journal of Computational and Graphical Statistics}
\bvolume{20}
\bpages{140--157}.
\end{barticle}
\endbibitem

\bibitem[\protect\citeauthoryear{Martens and Van~Dijk}{2007}]{Martens_2007}
\begin{barticle}[author]
\bauthor{\bsnm{Martens},~\bfnm{Martin}\binits{M.}} \AND
  \bauthor{\bsnm{Van~Dijk},~\bfnm{Dick}\binits{D.}}
(\byear{2007}).
\btitle{{Measuring Volatility with the Realized Range}}.
\bjournal{Journal of Econometrics}
\bvolume{138}
\bpages{181--207}.
\end{barticle}
\endbibitem

\bibitem[\protect\citeauthoryear{Moghadam and Vi{\~n}als}{2010}]{Moghadam_2010}
\begin{btechreport}[author]
\bauthor{\bsnm{Moghadam},~\bfnm{Reza}\binits{R.}} \AND
  \bauthor{\bsnm{Vi{\~n}als},~\bfnm{Jos{\'e}}\binits{J.}}
(\byear{2010}).
\btitle{{Understanding Financial Interconnectedness}}
\btype{Mimeo},
\bpublisher{International Monetary Fund}.
\end{btechreport}
\endbibitem

\bibitem[\protect\citeauthoryear{Sarkar and Moore}{2005}]{Sarkar_2005}
\begin{barticle}[author]
\bauthor{\bsnm{Sarkar},~\bfnm{Purnamrita}\binits{P.}} \AND
  \bauthor{\bsnm{Moore},~\bfnm{Andrew~W}\binits{A.~W.}}
(\byear{2005}).
\btitle{{Dynamic Social Network Analysis Using Latent Space Models}}.
\bjournal{ACM SIGKDD Explorations Newsletter}
\bvolume{7}
\bpages{31--40}.
\end{barticle}
\endbibitem

\bibitem[\protect\citeauthoryear{Sewell and Chen}{2015}]{Sewell_2015}
\begin{barticle}[author]
\bauthor{\bsnm{Sewell},~\bfnm{Daniel~K}\binits{D.~K.}} \AND
  \bauthor{\bsnm{Chen},~\bfnm{Yuguo}\binits{Y.}}
(\byear{2015}).
\btitle{{Latent Space Models for Dynamic Networks}}.
\bjournal{Journal of the American Statistical Association}
\bvolume{110}
\bpages{1646-1657}.
\end{barticle}
\endbibitem

\bibitem[\protect\citeauthoryear{Silva and Ghahramani}{2009}]{Silva_2009}
\begin{barticle}[author]
\bauthor{\bsnm{Silva},~\bfnm{Ricardo}\binits{R.}} \AND
  \bauthor{\bsnm{Ghahramani},~\bfnm{Zoubin}\binits{Z.}}
(\byear{2009}).
\btitle{{The Hidden Life of Latent Variables: Bayesian Learning with Mixed
  Graph Models}}.
\bjournal{Journal of Machine Learning Research}
\bvolume{10}
\bpages{1187--1238}.
\end{barticle}
\endbibitem

\bibitem[\protect\citeauthoryear{Tang et~al.}{2010}]{Tang_2010}
\begin{btechreport}[author]
\bauthor{\bsnm{Tang},~\bfnm{Chrismin}\binits{C.}},
  \bauthor{\bsnm{Dungey},~\bfnm{Mr~Mardi}\binits{M.~M.}},
  \bauthor{\bsnm{Martin},~\bfnm{Mr~Vance}\binits{M.~V.}},
  \bauthor{\bsnm{Gonz{\'a}lez-Hermosillo},~\bfnm{Ms~Brenda}\binits{M.~B.}} \AND
  \bauthor{\bsnm{Fry},~\bfnm{Ms~Renee}\binits{M.~R.}}
(\byear{2010}).
\btitle{{Are Financial Crises Alike?}}
\btype{Working Paper} No. \bnumber{10--14},
\bpublisher{International Monetary Fund}.
\end{btechreport}
\endbibitem

\bibitem[\protect\citeauthoryear{Tibshirani}{1996}]{Tibshirani_1996}
\begin{barticle}[author]
\bauthor{\bsnm{Tibshirani},~\bfnm{Robert}\binits{R.}}
(\byear{1996}).
\btitle{{Regression Shrinkage and Selection via the LASSO}}.
\bjournal{Journal of the Royal Statistical Society. Series B}
\bvolume{58}
\bpages{267--288}.
\end{barticle}
\endbibitem

\bibitem[\protect\citeauthoryear{Wang}{2015}]{Wang_2015}
\begin{barticle}[author]
\bauthor{\bsnm{Wang},~\bfnm{Hao}\binits{H.}}
(\byear{2015}).
\btitle{{Scaling It Up: Stochastic Search Structure Learning in Graphical
  Models}}.
\bjournal{Bayesian Analysis}
\bvolume{10}
\bpages{351--377}.
\end{barticle}
\endbibitem

\bibitem[\protect\citeauthoryear{Wang and Li}{2012}]{Wang_2012}
\begin{barticle}[author]
\bauthor{\bsnm{Wang},~\bfnm{Hao}\binits{H.}} \AND
  \bauthor{\bsnm{Li},~\bfnm{Sophia~Zhengzi}\binits{S.~Z.}}
(\byear{2012}).
\btitle{{Efficient Gaussian Graphical Model Determination Under G-Wishart Prior
  Distributions}}.
\bjournal{Electronic Journal of Statistics}
\bvolume{6}
\bpages{168--198}.
\end{barticle}
\endbibitem

\bibitem[\protect\citeauthoryear{Wang et~al.}{2010}]{Wang_2010}
\begin{barticle}[author]
\bauthor{\bsnm{Wang},~\bfnm{Chaolong}\binits{C.}},
  \bauthor{\bsnm{Szpiech},~\bfnm{Zachary~A}\binits{Z.~A.}},
  \bauthor{\bsnm{Degnan},~\bfnm{James~H}\binits{J.~H.}},
  \bauthor{\bsnm{Jakobsson},~\bfnm{Mattias}\binits{M.}},
  \bauthor{\bsnm{Pemberton},~\bfnm{Trevor~J}\binits{T.~J.}},
  \bauthor{\bsnm{Hardy},~\bfnm{John~A}\binits{J.~A.}},
  \bauthor{\bsnm{Singleton},~\bfnm{Andrew~B}\binits{A.~B.}} \AND
  \bauthor{\bsnm{Rosenberg},~\bfnm{Noah~A}\binits{N.~A.}}
(\byear{2010}).
\btitle{{Comparing Spatial Maps of Human Population-Genetic Variation Using
  Procrustes Analysis}}.
\bjournal{Statistical Applications in Genetics and Molecular Biology}
\bvolume{9}
\bpages{13}.
\end{barticle}
\endbibitem

\end{thebibliography}

\appendix
\section{Details of Sampling Approach of the Parameters}
\label{app}

This section provides a detailed descriptions of the sampling approach of the parameters.

\subsection{Sampling \texorpdfstring{$\Sigma$}{\Sigma}}

For $i=1,\ldots, n$ and $-i=\{1,\ldots, n\} \backslash \{i\},$ we partition $\Sigma$, $S_{y|x}$ and $V$ as follows
\begin{align}\label{partitions}
 \Sigma = \begin{pmatrix}
           \Sigma_{-i} & \sigma_{-i}\\
           \sigma_{-i}' & \sigma_{ii}
          \end{pmatrix}, \quad 
 S_{y|x} = \begin{pmatrix}
           S_{-i} & s_{-i}\\
           s_{-i}' & s_{ii}
          \end{pmatrix}, \quad 
 V = \begin{pmatrix}
           V_{-i} & v_{-i}\\
           v_{-i}' & v_{ii}
          \end{pmatrix}
\end{align}
where $\sigma_{ii}$, $s_{ii}$ and $v_{ii}$ are the $i$-th diagonal elements 
of $\Sigma$, $S_{y|x}$ and $V$ respectively, $\sigma_{-i}, s_{-i}$ and 
$v_{-i}$ are $(n-1)\times 1$ vectors, i.e., the rest of the elements on the 
$i$-th column of $\Sigma$, $S_{y|x}$ and $V$, and $\Sigma_{-i}, S_{-i}$ and $V_{-i}$ are $(n-1)\times (n-1)$ matrices. 

From the marginal likelihood function $P(Y|\Sigma)$ in (\ref{marginal likelihood}) and the priors $P(\Sigma|G)$ in (\ref{Prior on Sigma}) 
and $P(G|Z)$ in (\ref{Prior on G}), we obtain the following expression
 \begin{align}\label{posterior of Sigma}
 P(\Sigma| Y, G) ~& \propto~ |\Sigma|^{-\frac{T}{2}} \text{etr} \Big ( -\frac{1}{2}  \Sigma^{-1} S_{y|x} \Big ) 
 \prod_{i \not= j} \exp \Big ( -\frac{1}{2} V_{ij}^{-1}\Sigma_{i,j}^2 \Big ) 
 \prod_{i=1}^n \exp \Big (-\frac{1}{2} V_{ii} \Sigma_{i,i} \Big ) \enskip .
\end{align}
Using the Sherman-Morrison-Woodbury formula, the inverse 
of the partitioned $\Sigma$ is given by
\begin{align}\label{Sigma inverse}
\Sigma^{-1} = \begin{pmatrix} \Sigma_{-i}^{-1} + \Sigma_{-i}^{-1} \sigma_{-i} 
\gamma^{-1} \sigma_{-i}'\Sigma_{-i}^{-1}  & - \Sigma_{-i}^{-1} \sigma_{-i}\gamma^{-1} \\
-\sigma_{-i}'\Sigma_{-i}^{-1}\gamma^{-1} & \gamma^{-1} \end{pmatrix}
\end{align}
where $\gamma = \sigma_{ii} - \sigma_{-i}' \Sigma_{-i}^{-1}\sigma_{-i}$. 
The determinant of the partitioned $\Sigma$ is 
\begin{align}\label{determinant Sigma}
|\Sigma| = |\sigma_{ii} - \sigma_{-i}' \Sigma_{-i}^{-1}\sigma_{-i}| |\Sigma_{-i}| 
 = \gamma |\Sigma_{-i}| \enskip .
\end{align}
Following \cite{Wang_2015}, we consider block updates of $\Sigma$ by focusing on a column and row at a time.
From (\ref{posterior of Sigma}) and the partitions in (\ref{partitions}), the 
distribution of the elements of the $i$-th column in $\Sigma$, i.e. $(\sigma_{-i}, \sigma_{ii})$, 
conditional on $Y, \Sigma_{-i}, G$ is given by
\begin{align}
P(\sigma_{-i}, \sigma_{ii}| Y, \Sigma_{-i}, G) ~ \propto & ~
\gamma^{\frac{T}{2}} ~
\exp \Big(-\frac{1}{2} \Big[ ~ \sigma_{-i}' \Sigma_{-i}^{-1} S_{-i} 
\Sigma_{-i}^{-1}\sigma_{-i}\gamma^{-1}  ~ - ~ 2s_{-i}'\Sigma_{-i}^{-1}\sigma_{-i} \gamma^{-1}
\nonumber \\
  &~ + ~s_{ii} \gamma^{-1} ~ + ~ \sigma_{-i}' (v_{ii} \Sigma_{-i}^{-1} + D_v^{-1})\sigma_{-i} ~ + ~ v_{ii}\gamma \Big] \Big)
\end{align}
where $D_v = \text{diag}(v_{-i})$. We consider a change of variable $(\sigma_{-i}, \sigma_{ii}) \to (\mu, \gamma)$, where $\mu = \sigma_{-i}$ and  
$\gamma = \sigma_{ii} - \sigma_{-i}' \Sigma_{-i}^{-1}\sigma_{-i})$. The associated Jacobian is a constant independent of $(\mu, \gamma)$. 
Following Proposition 2 of \cite{Wang_2015}, the conditional distribution of $\mu$ and $\gamma$ given $Y, \Sigma_{-i}, G$ 
 is a Gaussian-generalized inverse Gaussian distribution such that 
 \begin{align*}
  \mu ~|~ Y, \Sigma_{-i}, G, \gamma & ~ \sim ~ \mathcal{N}\bigg( W^{-1}\Sigma_{-i}^{-1}s_{-i}\gamma^{-1}, ~W^{-1}   \bigg ) \\
  \gamma ~|~ Y, \Sigma_{-i}, G, \mu & ~ \sim ~ GIG \big(  q, a, b \big )
 \end{align*}
where $W=\Sigma_{-i}^{-1} S_{-i} \Sigma_{-i}^{-1} \gamma^{-1} + v_{ii} \Sigma_{-i}^{-1}+ D_v^{-1}$, and $(q, a, b)$ are the parameters of the 
generalized inverse Gaussian distribution (GIG), where $q=1-\frac{1}{2}T$, 
$~a=v_{ii}$, and $~b=\mu'\Sigma_{-i}^{-1} S_{-i}\Sigma_{-i}^{-1}\mu - 2s_{-i}'\Sigma_{-i}^{-1}\mu + s_{ii}$. The density of the GIG is given by 
\begin{align}
P(x|q,a,b) =\Big(\frac{a}{b}\Big)^{q/2}\frac{x^{q-1}}{2K_q(\sqrt{ab})}\exp \Big(-\frac{1}{2} \big[ax + b/x \big] \Big)
\end{align}
where $K_q$ is the modified Bessel function of the second kind.

\subsection{Sampling \texorpdfstring{$G$}{G}}

Combining $P(\Sigma|G)$ in (\ref{Prior on Sigma}) and $P(G|U, \Lambda,
\theta)$ in (\ref{Prior on G}), the conditional distribution of each edge 
$G_{ij}$ given $\Sigma$, $U$, $\Lambda$, and $\theta$ is independent Bernoulli
distributed, 
\[
 G_{ij}|\Sigma, U, \Lambda, \theta ~ \sim ~\text{Ber}\Bigg(\frac{b_{ij1}}{b_{ij1} + b_{ij2}}\Bigg)
\]
where $b_{ij1}  = \Gamma_{ij} / v_1 \exp\{-\sigma_{ij}^2/(2v_1^2)\}$, and
$b_{ij2} = (1 - \Gamma_{ij}) / v_0 \exp\{-\sigma_{ij}^2/(2v_0^2)\}$.

\subsection{Sampling \texorpdfstring{$Z$}{Z}}
Since $Z_{ij} | U, \Lambda, \theta \sim \mathcal{N}(\theta + (U\Lambda
U')_{ij}, 1)$ independently and $G_{ij} = \vect{1}(Z_{ij} > 0)$, we have:
\begin{align}\label{Pdf Z}
 Z_{ij} | G, U, \Lambda, \theta ~ \sim ~ \mathcal{N}(\theta + (U\Lambda U')_{ij}, 1) ~\vect{1}(Z_{ij} > 0)^{G_{ij}} ~\vect{1}(Z_{ij} < 0)^{1 - G_{ij}}
\end{align}
that is, each $Z_{ij}$ is independently distributed as a truncated version of
the prior but conditional on being positive or negatively truncated given
$G_{ij}$.

\subsection{Sampling \texorpdfstring{$\theta$}{\theta}}
Following the distribution of $Z$ in (\ref{Pdf Z}) and a normal prior distribution on $\theta$, the conditional distribution of 
$\theta$ given $\{Z, U, \Lambda\}$ is as follows:
\begin{align}\label{Posterior theta}
P(\theta| Z, U, \Lambda) ~ & \propto ~ \hbox{etr} \Big ( - \frac{1}{4} E_z'E_z \Big) ~ \hbox{etr} \Big ( ~ \frac{1}{2} E_z'U\Lambda U' \Big) ~ \exp 
\Big ( -\frac{1}{2\tau_{\theta}^2} (\theta - \theta_0)^2 \Big) \enskip .
\end{align}
The posterior distribution of 
\begin{align*}
 \theta| Z, U, \Lambda ~\sim ~ \mathcal{N} \bigg( ~\frac{2\tau_{\theta}^2}{2 + n(n-1)\tau_{\theta}^2} \Big(\sum_{j>i}(Z - U\Lambda U')_{ij} + \frac{\theta_0}{\tau_{\theta}^2} \Big), 
~\frac{2\tau_{\theta}^2}{2 + n(n-1)\tau_{\theta}^2} \bigg )
\end{align*}
where $\displaystyle \sum_{j>i}(Z - U\Lambda U')_{ij}$ is a summation of the upper off-diagonals of $(Z - U\Lambda U')$.

\subsection{Sampling \texorpdfstring{$\Lambda$}{\Lambda}}

Combining the distribution of $Z$ in (\ref{Pdf Z}) and the prior distribution of $\Lambda = \text{diag}(\lambda_1, \lambda_2)$ in (\ref{Prior lambda}), the conditional distribution of 
$\Lambda$ given $\{Z, \theta, U\}$ is as follows:
\begin{align}\label{Posterior Lambda}
P(\Lambda| Z, \theta, U) ~ & \propto ~ \hbox{etr} \Big ( - \frac{1}{2} \Big[  \frac{1}{2}\Lambda^2  -E_z'U\Lambda U' \Big]\Big)
\prod_{r=1}^2 \exp \Big( -\frac{1}{2} \frac{\lambda_r^2}{\tau_{\lambda}^2} \Big)
\enskip .
\end{align}
Let $U_r$ be the $r$-th column of $U$. The posterior distribution for $\lambda_r, ~r=1,2$ is given by 
\begin{align*}
 \lambda_r | Z, \theta, U ~ \sim ~ \mathcal{N}\bigg( ~\frac{\tau_{\lambda}^2}{2+\tau_{\lambda}^2} U_r'E_zU_r, ~\frac{2\tau_{\lambda}^2}{2+\tau_{\lambda}^2} \bigg )
\enskip .
\end{align*} 

\subsection{Sampling \texorpdfstring{$U$}{U}}

Following the literature on directional statistics, 
a random matrix $U$ distributed on $\mathcal{V}_r(\mathcal{R}^n)$ is said to have a matrix Bingham vMF distribution with density function given by 
 \begin{align}\label{Conditional Dist of U}
 P (U|H_a, H_b) ~\propto~ \hbox{etr} \Big(H_b U'H_aU \Big)
\end{align}
where $H_a$ is an $n\times n$ symmetric matrix and $H_b$ is an $r\times r$ diagonal matrix. Following the distribution of $Z$ in (\ref{Pdf Z}) and a uniform prior distribution on $U$, the 
conditional distribution involving $U$ given $\{Z, \theta, \Lambda\} $ is as follows:
\begin{align}\label{Posterior U}
P(U| \theta, Z, \Lambda) \approx P(U| E_z, \Lambda) ~ & \propto ~ \hbox{etr} \Big ( ~ \frac{1}{2} E_z'U\Lambda U' \Big) ~ = ~ \hbox{etr} \Big ( ~ \frac{1}{2} \Lambda U' E_z U \Big) \enskip .
\end{align}
Comparing (\ref{Posterior U}) and (\ref{Conditional Dist of U}), we set $H_a = E_z/2$ and $H_b=\Lambda$. This 
 corresponds to the matrix Bingham vMF (BvMF) distribution and the posterior of $U$ with respect to its uniform prior is $U| \theta, Z, \Lambda \sim BvMF (E_z/2, \Lambda)$. 

We sample the columns of $U$ by adopting the Gibbs approach in \cite{Hoff_2009} and \cite{Brubaker_2012}. 
Let $U_{-j} = U\backslash U_j$ denote $U$ excluding the $j$-th column. For a random draw of $r \sim \{1, 2 \}$, 
we perform the following: 
\begin{enumerate}
\item obtain $N_{-r}$,  the null space of $U_{-r}$ and compute $x = N_{-r}' U_r $. 
Note that the matrix $N_{-r} = null(U_{-r})$ such that $N_{-r}U_{-r}=0$  
\item compute $\tilde{H}_a = H_{b(r,r)}N_{-r}'H_aN_{-r}$, where $H_{b(r,r)}$ is the $r$-th row and $r$-th column of $H_b$ 
\item update $x\sim P(x|\tilde{H}_a) \propto \exp (x'\tilde{H}_a x)$ following \citep{Brubaker_2012}
\item set $U_r = N_ {-r}x$
\end{enumerate}

\section{Financial Data Description}

\begin{center}
\scriptsize
\begin{longtable}{@{\extracolsep{\fill}}l l l l l@{}}
\caption{Financial Data Description Classified By Country and Industry.}
\label{Dataset 1}
\\
\toprule
No. & Institution & Ticker  & Country/Region & Industry  \\
\toprule
\endfirsthead
\caption[]{Financial Data Description \textit{(Continuation)}}\\
\toprule
No. & Institution & Ticker  & Country/Region & Industry  \\
\toprule
\endhead
\bottomrule 
\multicolumn{5}{r}{\textit{Continued on next page}} \\
\endfoot
\bottomrule 
\endlastfoot
 1 &  S\&P 500 & GSPC.US & North America & Market Index \\ 
 2 &  Dow Jones & DJI.US & North America & Market Index \\ 
 3 &  Euro Stoxx 600 & DJSTOXX.EU & Europe & Market Index \\ 
 4 &  Hang Seng Index & HSI.HK & Asia & Market Index \\ 
 5 &  Nasdag Composite & IXIC.US & North America & Market Index \\ 
 6 &  Euro Stoxx 50 & STOXX50E.EU & Europe & Market Index \\ \midrule
 7 &  Immofiz & IIA.AT & Austria  & Real Estate \\ 
 8 &  Vienna Insurance Group A & VIG.AT & Austria  & Insurance \\ 
 9 &  Cofinimmo & COFB.BE &  Belgium  & Real Estate \\ 
 10 &  Credit Suisse Group N & CSGN.CH &  Switzerland  & Bank \\ 
 11 &  Helvetia Holding N & HELN.CH &  Switzerland  & Insurance \\ 
 12 &  PSP Swiss Property Ag & PSPN.CH &  Switzerland  & Real Estate \\ 
 13 &  Swiss Life Holding & SLHN.CH &  Switzerland  & Insurance \\ 
 14 &  Swiss Prime Site & SPSN.CH &  Switzerland  & Real Estate \\ 
 15 &  Swiss Re & SREN.CH &  Switzerland  & Insurance \\ 
 16 &  UBS & UBSN.CH &  Switzerland  & Bank \\ 
 17 &  Zurich Insurance Group & ZURN.CH &  Switzerland  & Insurance \\ 
 18 &  Allianz  & ALV.DE & Germany  & Insurance \\ 
 19 &  Commerzbank  & CBK.DE & Germany  & Bank \\ 
 20 &  Deutsche Bank  & DBK.DE & Germany  & Bank \\ 
 21 &  Hannover Ruck.  & HNR.DE & Germany  & Insurance \\ 
 22 &  Muenchener Rueck.  & MUV.DE & Germany  & Insurance \\ 
 23 &  Danske Bank & DANSKE.DK &  Denmark  & Bank \\ 
 24 &  Topdanmark & TOP.DK &  Denmark  & Insurance \\ 
 25 &  BBV Argentaria & BBVA.ES &  Spain  & Bank \\ 
 26 &  Mapfre & MAP.ES &  Spain  & Insurance \\ 
 27 &  Banco Santander & SAN.ES &  Spain  & Bank \\ 
 28 &  Sampo A & SAMPO.FI &  Finland  & Insurance \\ 
 29 &  Credit Agricole & ACA.FR &  France  & Bank \\ 
 30 &  BNP Paribas & BNP.FR &  France  & Bank \\ 
 31 &  CNP Assurances & CNP.FR &  France  & Insurance \\ 
 32 &  AXA & CS.FR &  France  & Insurance \\ 
 33 &  Fonciere Des Regions & FDR.FR &  France  & Real Estate \\ 
 34 &  Gecina & GFC.FR &  France  & Real Estate \\ 
 35 &  Societe Generale & GLE.FR &  France  & Bank \\ 
 36 &  ICADE & ICAD.FR &  France  & Real Estate \\ 
 37 &  Natixis & KN.FR &  France  & Bank \\ 
 38 &  Klepierre & LI.FR &  France  & Real Estate \\ 
 39 &  Scor SE & SCR.FR &  France  & Insurance \\ 
 40 &  National Bank of Greece & ETE.GR &  Greece  & Bank \\ 
 41 &  Piraeus Bank & TPEIR.GR &  Greece  & Bank \\ 
 42 &  Bank of Ireland & BIR.IR &  Ireland  & Bank \\ 
 43 &  Banca Monte Dei Paschi & BMPS.IT &  Italy  & Bank \\ 
 44 &  Assicurazioni Generali & G.IT &  Italy  & Insurance \\ 
 45 &  Intesa Sanpaolo & ISP.IT &  Italy  & Bank \\ 
 46 &  Unicredit & UCG.IT &  Italy  & Bank \\ 
 47 &  Unipolsai & US.IT &  Italy  & Insurance \\ 
 48 &  Aegon & AGN.NL & Netherland  & Insurance \\ 
 49 &  ING Groep & INGA.NL & Netherland  & Bank \\ 
 50 &  Unibail-Rodamco & UL.NL & Netherland  & Real Estate \\ 
 51 &  Wereldhave & WHA.NL & Netherland  & Real Estate \\ 
 52 &  DNB & DNB.NO &  Norway  & Bank \\ 
 53 &  Storebrand & STB.NO &  Norway  & Insurance \\ 
 54 &  Castellum & CAST.SE &  Sweden  & Real Estate \\ 
 55 &  JM & JM.SE &  Sweden  & Real Estate \\ 
 56 &  Nordea Bank & NDA.SE &  Sweden  & Bank \\ 
 57 &  Aviva & AV.UK & United Kingdom & Insurance \\ 
 58 &  Barclays & BARC.UK & United Kingdom & Bank \\ 
 59 &  British Land & BLND.UK & United Kingdom & Real Estate \\ 
 60 &  Derwent London & DLN.UK & United Kingdom & Real Estate \\ 
 61 &  Great Portland Estates & GPOR.UK & United Kingdom & Real Estate \\ 
 62 &  Hammerson & HMSO.UK & United Kingdom & Real Estate \\ 
 63 &  HSBC Hdg. & HSBA.UK & United Kingdom & Bank \\ 
 64 &  Hiscox & HSX.UK & United Kingdom & Insurance \\ 
 65 &  Intu Properties & INTU.UK & United Kingdom & Real Estate \\ 
 66 &  Land Securities Group & LAND.UK & United Kingdom & Real Estate \\ 
 67 &  Legal \& General & LGEN.UK & United Kingdom & Insurance \\ 
 68 &  Lloyds Banking Group & LLOY.UK & United Kingdom & Bank \\ 
 69 &  Old Mutual & OML.UK & United Kingdom & Insurance \\ 
 70 &  Prudential & PRU.UK & United Kingdom & Insurance \\ 
 71 &  Royal Bank Of Sctl.Gp. & RBS.UK & United Kingdom & Bank \\ 
 72 &  RSA Insurance Group & RSA.UK & United Kingdom & Insurance \\ 
 73 &  Segro & SGRO.UK & United Kingdom & Real Estate \\ 
 74 &  Shaftesbury & SHB.UK & United Kingdom & Real Estate \\ 
 75 &  Standard Chartered & STAN.UK & United Kingdom & Bank \\  
 76 &  St.Jamess Place & STJ.UK & United Kingdom & Insurance \\ 
 77 &  Aflac & AFL.US & United States & Insurance \\ 
 78 &  American Intl. Gp. & AIG.US & United States & Insurance \\ 
 79 &  Arthur J Gallagher & AJG.US & United States & Insurance \\ 
 80 &  Allstate & ALL.US & United States & Insurance \\ 
 81 &  AON Class A & AON.US & United States & Insurance \\ 
 82 &  American Express & AXP.US & United States & Bank \\ 
 83 &  Bank of America & BAC.US & United States & Bank \\ 
 84 &  BB\&T & BBT.US & United States & Bank \\ 
 85 &  Bank of New York Mellon & BK.US & United States & Bank \\ 
 86 &  BOK Finl. & BOKF.US & United States & Bank \\ 
 87 &  Berkshire Hathaway A & BRKA.US & United States & Insurance \\ 
 88 &  Brown \& Brown & BRO.US & United States & Insurance \\ 
 89 &  Citigroup & C.US & United States & Bank \\ 
 90 &  Chubb & CB.US & United States & Insurance \\ 
 91 &  Comerica & CMA.US & United States & Bank \\ 
 92 &  CNA Financial & CNA.US & United States & Insurance \\ 
 93 &  Capital One Finl. & COF.US & United States & Bank \\ 
 94 &  Corrections Amer New & CXW.US & United States & Real Estate \\ 
 95 &  Duke Realty Corporation & DRE.US & United States & Real Estate \\ 
 96 &  Equity Lifestyle Props. & ELS.US & United States & Real Estate \\ 
 97 &  Essex Property Tst. & ESS.US & United States & Real Estate \\ 
 98 &  Fifth Third Bancorp & FITB.US & United States & Bank \\ 
 99 &  Federal Realty Inv.Tst. & FRT.US & United States & Real Estate \\ 
 100 &  General Gw.Props. & GGP.US & United States & Real Estate \\ 
 101 &  Goldman Sachs Gp. & GS.US & United States & Bank \\ 
 102 &  Huntington Bcsh. & HBAN.US & United States & Bank \\ 
 103 &  Hudson City Banc. & HCBK.US & United States & Bank \\ 
 104 &  HCC Insurance Hdg. & HCC.US & United States & Insurance \\ 
 105 &  Welltower & HCN.US & United States & Real Estate \\ 
 106 &  HCP & HCP.US & United States & Real Estate \\ 
 107 &  Hartford Finl.Svs.Gp. & HIG.US & United States & Insurance \\ 
 108 &  Host Hotels \& Resorts & HST.US & United States & Real Estate \\ 
 109 &  JP Morgan & JPM.US & United States & Bank \\ 
 110 &  Keycorp & KEY.US & United States & Bank \\ 
 111 &  Kimco Realty & KIM.US & United States & Real Estate \\ 
 112 &  Loews & L.US & United States & Insurance \\ 
 113 &  Lincoln National & LNC.US & United States & Insurance \\ 
 114 &  Liberty Property Tst. & LPT.US & United States & Real Estate \\ 
 115 &  Macerich & MAC.US & United States & Real Estate \\ 
 116 &  Metlife & MET.US & United States & Insurance \\ 
 117 &  Markel & MKL.US & United States & Insurance \\ 
 118 &  Marsh \& Mclen & MMC.US & United States & Insurance \\ 
 119 &  Morgan Stanley & MS.US & United States & Bank \\ 
 120 &  M\&T Bank & MTB.US & United States & Bank \\ 
 121 &  Mitsubishi UFJ Fin.Gp & MTU.US & United States & Insurance \\ 
 122 &  Northern Trust & NTRS.US & United States & Bank \\ 
 123 &  Realty Income & O.US & United States & Real Estate \\ 
 124 &  Principal Finl.Gp. & PFG.US & United States & Insurance \\ 
 125 &  Progressive Ohio & PGR.US & United States & Insurance \\ 
 126 &  Prologis & PLD.US & United States & Real Estate \\ 
 127 &  PNC Finl.Svs.Gp. & PNC.US & United States & Bank \\ 
 128 &  Prudential Finl. & PRU.US & United States & Insurance \\ 
 129 &  Public Storage & PSA.US & United States & Real Estate \\ 
 130 &  Regency Centers & REG.US & United States & Real Estate \\ 
 131 &  Regions Finl.New & RF.US & United States & Bank \\ 
 132 &  Charles Schwab & SCHW.US & United States & Bank \\ 
 133 &  Sl Green Realty & SLG.US & United States & Real Estate \\ 
 134 &  Simon Property Group & SPG.US & United States & Real Estate \\ 
 135 &  Suntrust Banks & STI.US & United States & Bank \\ 
 136 &  State Street & STT.US & United States & Bank \\ 
 137 &  Torchmark & TMK.US & United States & Insurance \\ 
 138 &  Travelers Cos. & TRV.US & United States & Insurance \\ 
 139 &  Unum Group & UNM.US & United States & Insurance \\ 
 140 &  US Bancorp & USB.US & United States & Bank \\ 
 141 &  Vornado Realty Trust & VNO.US & United States & Real Estate \\ 
 142 &  Ventas & VTR.US & United States & Real Estate \\ 
 143 &  Wells Fargo \& Co & WFC.US & United States & Bank \\ 
 144 &  W R Berkley & WRB.US & United States & Insurance \\ 
 145 &  Alleghany & Y.US & United States & Insurance \\ 
 146 &  Zions Bancorp. & ZION.US & United States & Bank \\ \midrule
 147 &  Bear Stearns\footnotemark[1] & BSC.US & United States & Bank  \\ 
 148 &  Countrywide Financial Corp.\footnotemark[2] & CCR.US & United States & Bank \\ 
 149 &  Lehman Brothers\footnotemark[3] & LEHM.US & United States & Bank \\ 
 150 &  Merrill Lynch\footnotemark[4] & MER.US & United States & Bank  \\
 \end{longtable}
\footnotetext[1]{Acquired by JP Morgan Chase on 3/17/2008.}
\footnotetext[2]{Acquired by Bank of America on  7/1/2008.}
\footnotetext[3]{The fourth-largest US investment bank before declaring bankruptcy on 9/15/2008.}
\footnotetext[4]{The third-largest US investment bank acquired by Bank of America on  1/1/2009.}
\end{center}   

\end{document}